\documentclass[aps,prd, floatfix,twocolumn,preprintnumbers,superscriptaddress,showpacs]{revtex4-1}
\usepackage{graphicx}
\usepackage{slashed}
\usepackage{amsmath,amsfonts,bm,bbold}
\usepackage{amssymb}
\begin{document}

\preprint{JLAB-THY-11-06}

\author{A.~V.~Radyushkin}
\affiliation{Physics Department, Old Dominion University, Norfolk,
             VA 23529, USA}
\affiliation{Thomas Jefferson National Accelerator Facility,
              Newport News, VA 23606, USA}
\affiliation{Bogoliubov Laboratory of Theoretical Physics, JINR, Dubna, Russian
             Federation}

\title{Generalized Parton Distributions  and Their Singularities}  

\begin{abstract}

A new approach to building models of generalized parton  distributions (GPDs) 
is discussed that is based on  the 
factorized DD (double distribution) Ansatz within the single-DD formalism.
The latter was not used before, because  
reconstructing GPDs  from the forward 
limit   one should 
start  in this  case with a very  singular function  $f(\beta)/\beta$ 
rather than with  the usual parton density  $f(\beta)$.  
This  results in a non-integrable
singularity  at $\beta=0$ exaggerated by the fact that 
 $f(\beta)$'s, on their own,  have a  singular $\beta^{-a}$  Regge behavior 
 for small $\beta$.
 It is  shown that the singularity is regulated within
 the  GPD  model of Szczepaniak et al., 
 in which the Regge behavior is implanted 
 through a subtracted dispersion relation 
 for the hadron-parton scattering amplitude.
 It  is  demonstrated that using  proper  softening of  the 
 quark-hadron vertices in the regions of large 
 parton virtualities results in model GPDs $H(x,\xi)$ that are
 finite  and continuous  at the ``border point''  $x=\xi$.
 Using a simple 
 input forward distribution,
we illustrate
implementation of  the new approach  for explicit 
construction
of model GPDs.
 As a further development, a  more general
 method of regulating  the $\beta=0$ singularities
 is proposed that is based on the separation
 of the initial single DD $f(\beta, \alpha)$ into the 
``plus'' part $[f(\beta,\alpha)]_{+}$
and the $D$-term. 
It is demonstrated that the ``DD+D'' separation
method allows to (re)derive GPD sum rules
that relate the difference between the forward
distribution $f(x)=H(x,0)$ and the border function 
$H(x,x)$ with the $D$-term function $D(\alpha)$.

\end{abstract}
\pacs{11.10.-z,12.38.-t,13.60.Fz}
\maketitle

\section{Introduction}

The ongoing  and  future experimental  studies of Generalized Parton Distributions
(GPDs)  \cite{Mueller:1998fv,Ji:1996ek,Radyushkin:1996nd,Collins:1996fb}
require theoretical   models for  GPDs  which  satisfy several nontrivial requirements,
such as polynomiality \cite{Ji:1998pc},  positivity \cite{Martin:1997wy,Pire:1998nw,Radyushkin:1998es}
  hermiticity  \cite{Mueller:1998fv}, time reversal invariance \cite{Ji:1998pc}, etc.,   
following from the most  general principles  of  quantum  field  theory.
In  particular, the polynomiality requirement, which  states that 
the $x^n$ moment of a GPD $H(x,\xi;t)$ is  a polynomial in $\xi$ of the
order not  higher than $n+1$, is a consequence 
of the Lorentz  invariance. 
The polynomiality condition is  automatically satisfied
when GPDs are constructed from Double Distributions (DDs) 
 \cite{Mueller:1998fv,Radyushkin:1996nd,Radyushkin:1996ru,Radyushkin:1998es},
 (see also  \footnote{``Dual  parameterization''
 \cite{Polyakov:2002wz,Polyakov:2007rw,Polyakov:2007rv,SemenovTianShansky:2008mp,Polyakov:2008aa,SemenovTianShansky:2010zv}
 is another way to impose the 
 polynomiality   condition  onto model GPDs.}), 
 thus the problem of  constructing a  model  for a  GPD 
 converts into a  problem of  building a  model
 for the relevant DD $F(\beta, \alpha;t)$.
 
 Since a DD 
 $F(\beta, \alpha;t)$ has hybrid properties:  it behaves like 
 a usual parton distribution function (PDF)  with respect to $\beta$,  as a meson 
 distribution amplitude  (DA) with respect to $\alpha$, and as  a form  factor 
 with respect to the invariant  momentum transfer $t$,
 it was proposed \cite{Radyushkin:1998es,Radyushkin:1998bz} 
(in the simplified formal $t=0$  limit) to  build a  model DD
$F(\beta, \alpha)$ as  a product of the usual  PDF $f(\beta)$
and a  profile function $h(\beta, \alpha)$  that  has an $\alpha$-shape of a 
meson DA.   This construction allows one   to get an intuitive feeling
about the   shape of GPDs  and their  change with  the change 
of the skewness parameter $\xi$. It  was noticed  \cite{Polyakov:1999gs}, however, 
that in  the case of isosinglet GPDs,  such a  Factorized DD Ansatz (FDDA)
does not produce the highest, $(n+1)^{\rm st}$  power of $\xi$ in the $x^n$ moment of $H(x,\xi)$.
To cure this problem, a  ``two-DD''  parameterization was proposed
\cite{Polyakov:1999gs}, with the second DD $G(\beta, \alpha)$
capable of generating,  among others,  the required $\xi^{n+1}$ power. 
It was also proposed \cite{Polyakov:1999gs} to use  a  ``DD plus D'' parameterization
in which the second DD $G(\beta, \alpha)$  is reduced to a function $D(\alpha)$
of one variable,  the D-term , that is   solely  responsible for the $\xi^{n+1}$ contribution.
The importance of the $D$-term and  its  physical  interpretation
was studied in further works (see Ref.~\cite{Goeke:2001tz}  and references therein).

Later, it was  found out that it is still  possible to write 
a  ``single-DD'' parameterization \cite{Belitsky:2000vk}
that   incorporates just one function, but produces 
all the required powers up to $\xi^{n+1}$.
This  representation also  has a remarkable property that 
it allows, in  principle,  to invert the GPD/DD relation,
i.e., to obtain  DD if GPD is known. 
So far,  however, the single-DD  representation was not 
used for  building   models for GPDs  using the factorized DD
Ansatz.  The reason is that  one should use  much  more singular function  $f(\beta)/\beta$ 
rather than just the usual PDF $f(\beta)$ for the  GPD  reconstruction from the forward 
limit.  The combination $f(\beta)/\beta$, being an even function 
in the singlet case, has a  non-integrable
singularity  at $\beta=0$, even if $f(\beta)$ is   finite at $\beta=0$.
Furthermore, the fact that  PDFs $f(\beta)$ have a  singular $\beta^{-a}$  Regge behavior 
makes the problem  even worse. 

In an independent development \cite{Szczepaniak:2007af}, 
an attempt was made to implant the  Regge behavior into 
a GPD model  constructed in the spirit of the covariant 
parton model \cite{Landshoff:1970ff}, with the hadron-parton transition
amplitude written in the dispersion relation
representation capable of generating the  desired   $s^a$  Regge
behavior through an appropriately chosen
spectral density.  To  handle $a>0$, 
the  subtracted dispersion relation was used.
The outcome was the claim \cite{Szczepaniak:2007af}
that the GPDs  $H(x,\xi)$  in this model have a singular
$(x-\xi)^{-a} $ behavior in the vicinity of the
``border'' point $x=\xi$, which, if true, would ruin  the 
applicability of the perturbative QCD formalism 
employing GPDs, since the latter   works only when the 
GPDs are  finite and  continuous across the border point $x=\xi$.

Our starting goal was to examine the model of   
Ref.~\cite{Szczepaniak:2007af} and to pinpoint  the physical
assumptions that resulted in the prediction of the singular 
$(x-\xi)^{-a} $ behavior  (for an earlier analysis, see Ref.~\cite{Kumericki:2007mh}).
As our analysis shows, the singularity follows from the use
of  the point-like approximation for the hadron-parton vertices.
For bound states, however, one expects that
the hadron wave function would generate 
an additional  power-like (or even exponential)  suppression
in the regions where the parton virtuality $k^2$ is large.
We found, that 
if such a suppression is properly included  in the model, the resulting GPDs
are finite  and continuous  for $x=\xi$.

In our study, we  also observed that the 
expression  for  GPDs  derived from the  model
of Ref.~\cite{Szczepaniak:2007af} corresponds to a single-DD representation.
Moreover, it has the structure of a factorized DD Ansatz,
but with the singularity  at  $\beta=0$  regularized 
by the subtraction  made in the dispersion relation 
for the quark-hadron scattering amplitude.
Thus, the  model  of Ref.~\cite{Szczepaniak:2007af} 
(corrected for an appropriate softening of the 
hadron-parton vertices) gives a framework 
for building GPD models within the 
single-DD scheme.  
Using a simple, but rather realistic 
model for the input forward distribution (i.e.,  usual PDF),
we illustrate, step by step, how to use this framework for 
the construction
of GPDs.

In particular, we found that  the    model produces a \mbox{$D$-term}  contribution,
despite the fact that it uses only the forward distribution 
as an input. 
The formal reason is that 
the subtraction introduced in  the dispersion relation 
differs from the subtraction that converts
the original DD $f(\beta,\alpha)$
 into a (mathematical) ``plus'' distribution $[f(\beta,\alpha)]_{+}$, which, 
 by definition,  
 cannot generate a $D$-term. 
 This  observation 
 raises the questions of a general nature about 
 the separation of the $D$-term from the initial DD 
 $f(\beta, \alpha)$  of 
 the single-DD formalism. 
 
 We  found  that the  separation of $f(\beta, \alpha)$ into the 
``plus'' part $[f(\beta,\alpha)]_{+}$
and the $D$-term
can be used to rederive 
the sum rule \cite{Teryaev:2005uj,Anikin:2007yh,Kumericki:2007sa,Diehl:2007jb,Teryaev:2010zz} related to the dispersion
relation for the real part of the DVCS amplitude \cite{Teryaev:2005uj,Anikin:2007yh,Kumericki:2007sa,Diehl:2007jb,Teryaev:2010zz},
and we also gave the derivation 
of  another  sum rule  \cite{Anikin:2007yh} proposed as the $\xi \to 0$ 
limit of that  generic sum rule,
and which relates the difference between the forward
distribution $f(x)=H(x,0)$ and the border function 
$H(x,x)$ with the $D$-term function $D(\alpha)$.

The paper is organized as follows.
To make it self-contained, we start, in Sect. II,  with 
a short review of the basic facts about DDs and GPDs.
In Sect. III, we describe  the model \cite{Szczepaniak:2007af} with
implanted   Regge behavior, and give our
derivation of   expressions 
for GPDs and DDs that follow from this model.
We stress the necessity of a 
profile function that eliminates the singularities
for $x=\xi$ and present explicit results
for models with two simplest non-flat profiles.
In Sect. IV, we perform a model-independent study 
of GPD sum rules, using the procedure of separating the 
 initial  DD  into its ``plus'' part and the $D$-term.
We emphasize that $H(x,0)/x$ 
and $H(x,x)/x$, due to their singular nature,  should be treated 
as  (mathematical)  distributions  rather than functions.
Finally, we summarize the paper.

\section{Preliminaries}

\subsection{Double distributions}

\begin{figure}[t]
\begin{center} 
 \includegraphics[scale=0.15]{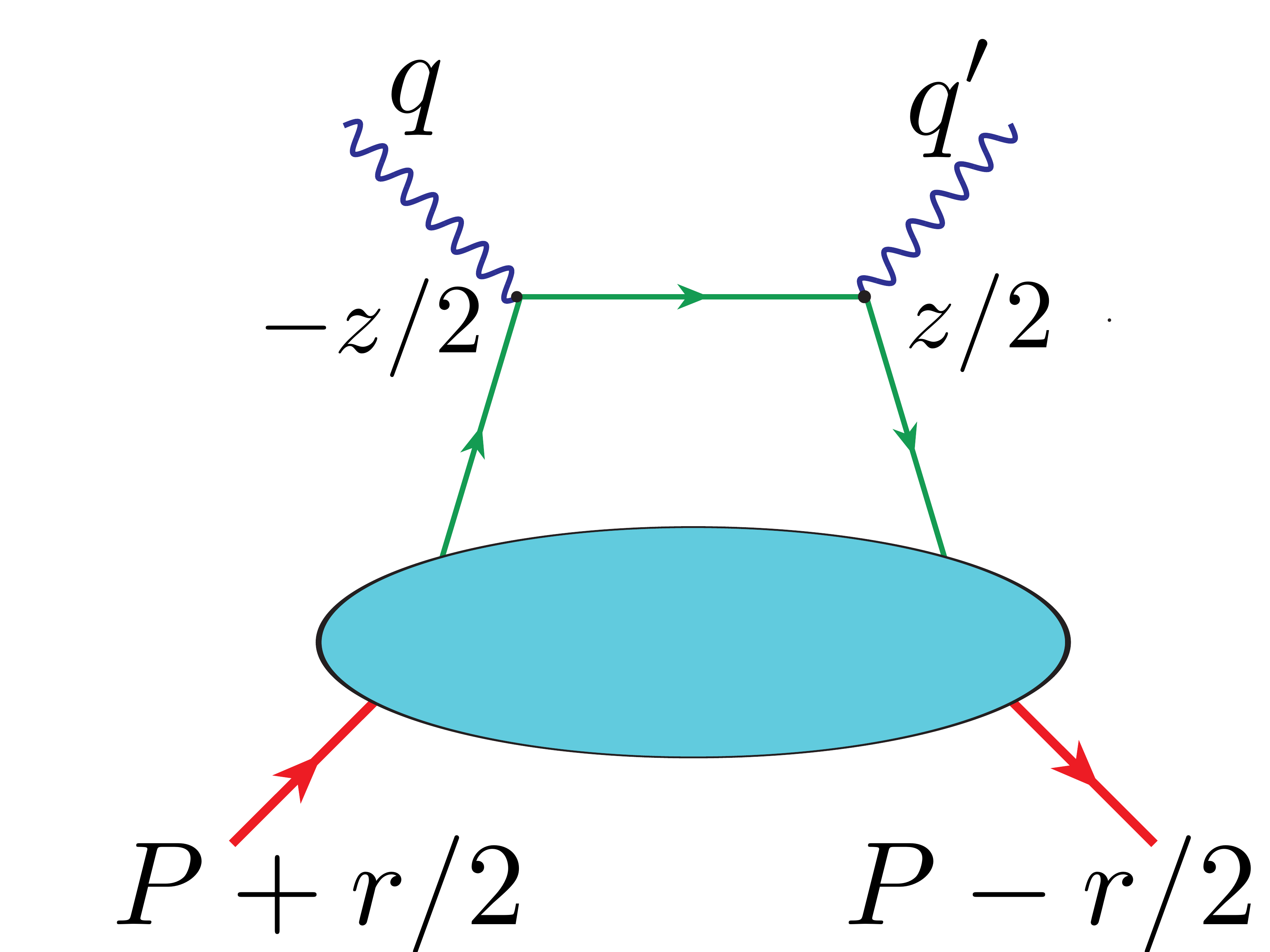} 
\end{center}
%\vspace{-5mm}
\caption{Structure of the handbag  diagram for deeply virtual
Compton scattering.}
\label{dvcs}
\end{figure}

Generalized parton distributions  (GPDs) \cite{Mueller:1998fv,Ji:1996ek,Radyushkin:1996nd,Collins:1996fb}
naturally appear in the perturbative QCD
description of the deeply virtual Compton scattering (DVCS) 
\cite{Ji:1996nm},\cite{Radyushkin:1996nd} (for reviews see 
\cite{Ji:1998pc, Radyushkin:2000uy, Goeke:2001tz, Diehl:2003ny, Belitsky:2005qn,Boffi:2007yc}),
the process in which
a highly virtual photon with momentum $q$,  upon scattering on a hadron 
 converts into  a real photon
with momentum $q'=q+r$.  
Basic features of GPD construction, in fact, 
are not specific to QCD, and may be illustrated 
on examples of  simpler theories \cite{Radyushkin:1998rt}.  
 In a toy scalar  model (scalar quarks $\psi$ 
 interacting with   a scalar photon   $\phi$  through $\psi\psi \phi$ vertex),
the  lowest-order  (handbag)  diagram (see Fig.\ref{dvcs}) may  be written,
in the coordinate representation,  as
\begin{align}
C(q,P,r) &= \int {\cal D}  (z)  e^{-i(qz)/2-i(q'z)/2} \nonumber \\
  \times & \langle P-r/2 |\psi (z/2) \psi(-z/2)|P+r/2 \rangle \, d^4z   \   ,
\label{Comp}
\end{align}
where $z$ is the separation between the ``photon'' vertices, and
$P=(p+p')/2$     is the average of the initial $p$ and the final $p'$
momenta of the struck hadron, and ${\cal D}(z)$ is the quark propagator.  

The matrix element may  depend
 on the coordinate difference $z$ 
through  invariants  $(Pz),(rz)$ and $z^2$  only.
For large $Q^2=-q^2$,  the higher terms of the $z^2$  expansion have  $1/Q^2$ 
suppression,  thus the leading power term is 
generated from the matrix element taken at  $z^2=0$.
The extraction of the $z^2=0$  part of the matrix element
may be performed  in the standard way:
through Taylor expansion in $z$ followed by  taking only  the 
symmetric-traceless part (denoted  by $\{  \  \}$)
$$ \psi(0) \{\stackrel{\leftrightarrow}{\partial}_{\mu_1} \ldots  \stackrel{\leftrightarrow}{\partial}_{\mu_n}\} \psi (0) 
$$
of the resulting local operators. 
For a scalar target,   one  may write
\begin{align}
&  \langle P+r/2 |   \psi(0) \{\stackrel{\leftrightarrow}{\partial}_{\mu_1} 
\ldots  \stackrel{\leftrightarrow}{\partial}_{\mu_n}\} \psi (0)|P-r/2 \rangle  \nonumber \\  &=
\sum_{n=0}^{\infty}\biggl [\sum_{l=0}^{n-1}A_{nl}\{P_{\mu_1}\ldots  P_{\mu_{n-l}}
r_{\mu_{n-l+1}}  \ldots   r_{\mu_n}\}\nonumber \\  &+  A_{nn}  \{r_{\mu_{1}}  \ldots   r_{\mu_n}\}  \biggr   ]  \  .
\label{scalarOn}
\end{align}
In the momentum  representation,   the  derivative $ \stackrel{\leftrightarrow}{\partial}_{\mu}\ $
converts  into  the   average $\bar k _\mu =(k_\mu+k'_\mu)/2$ of  the   initial  $k$  and final $k'$ quark  
momenta. After integration over $k$, $(\bar k)^n$ should produce the $P$  and $r$  factors
in the r.h.s.  of the equation above. In this   sense, one may treat $(\bar k)^n$
as $(\beta P + \alpha r/2)^n$  and   define the {\it double distribution}  (DD)
 \cite{Mueller:1998fv,Radyushkin:1996nd,Radyushkin:1996ru,Radyushkin:1998es} 
\begin{align}
 \frac{n!}{(n-l)! \,  l! \, 2^{l}} \int_{\Omega}  F(\beta, \alpha) \beta^{n-l} \alpha^l \, d\beta \, d\alpha=
A_{nl}
\end{align}
as a function whose $\beta^{n-l} \alpha^l$   moments are  proportional to the coefficients
$A_{nl}$.   It   can  be  shown  \cite{Mueller:1998fv,Radyushkin:1996nd,Radyushkin:1998bz} that the support region
$\Omega$  is given by the rhombus $|\alpha|+|\beta| \leq 1$.
These   definitions result in the \mbox{``DD   parameterization''} 
\begin{align}
& \langle P-r/2 |   \psi(-z/2) \psi (z/2)|P+r/2 \rangle  \nonumber \\  &
=  
\int_{\Omega}  F(\beta, \alpha) \,  e^{-i \beta (Pz) -i\alpha (rz)/2} \, d\beta \, d\alpha
 +  {\cal O} (z^2) \   . 
\label{DDF}
\end{align}
 of the
matrix element. 

\subsection{Generalized parton distributions}

Substituting the DD parameterization of the matrix element into the 
expression for the Compton amplitude, one obtains
\begin{align}
&  C(q,P,r) = 
\int_{\Omega}  F(\beta, \alpha) \,  D \left [q+\beta P +  \frac{1+\alpha}{2}r \right ]  \, d\beta \, d\alpha  \  ,
\label{Comptmom}
\end{align}
where $D [l]$   is the quark propagator in the   momentum  representation.
Thus, the leading-twist term  corresponds to a  parton  picture
in  which the initial quark  carries  momentum $\beta P + (1+\alpha) r/2$.
Neglecting   $P^2, (Pr)$  and $ r^2$,  we get  
$$(q+\beta P +  (1+\alpha)r/2)^2 = -Q^2 + 2\beta(Pq') + (1+\alpha) (rq') \  ,$$
i.e., $\beta$ and $\alpha$  appear in the propagator  in the combination
$\beta (Pq') +\alpha (rq')/2$  only. The latter  may  be written as $x (Pq')$,
with $x= \beta + \xi \alpha$, where $\xi = (rq')/2(Pq')$.
This redefinition leads to the parton  picture in which 
the initial quark  carries momentum $(x+\xi)P$. Introducing 
the {\it generalized parton distribution} (GPD)  \cite{Mueller:1998fv,Ji:1996ek,Radyushkin:1996ru} 
\begin{figure}[t]
\begin{center} 
 \includegraphics[scale=0.25]{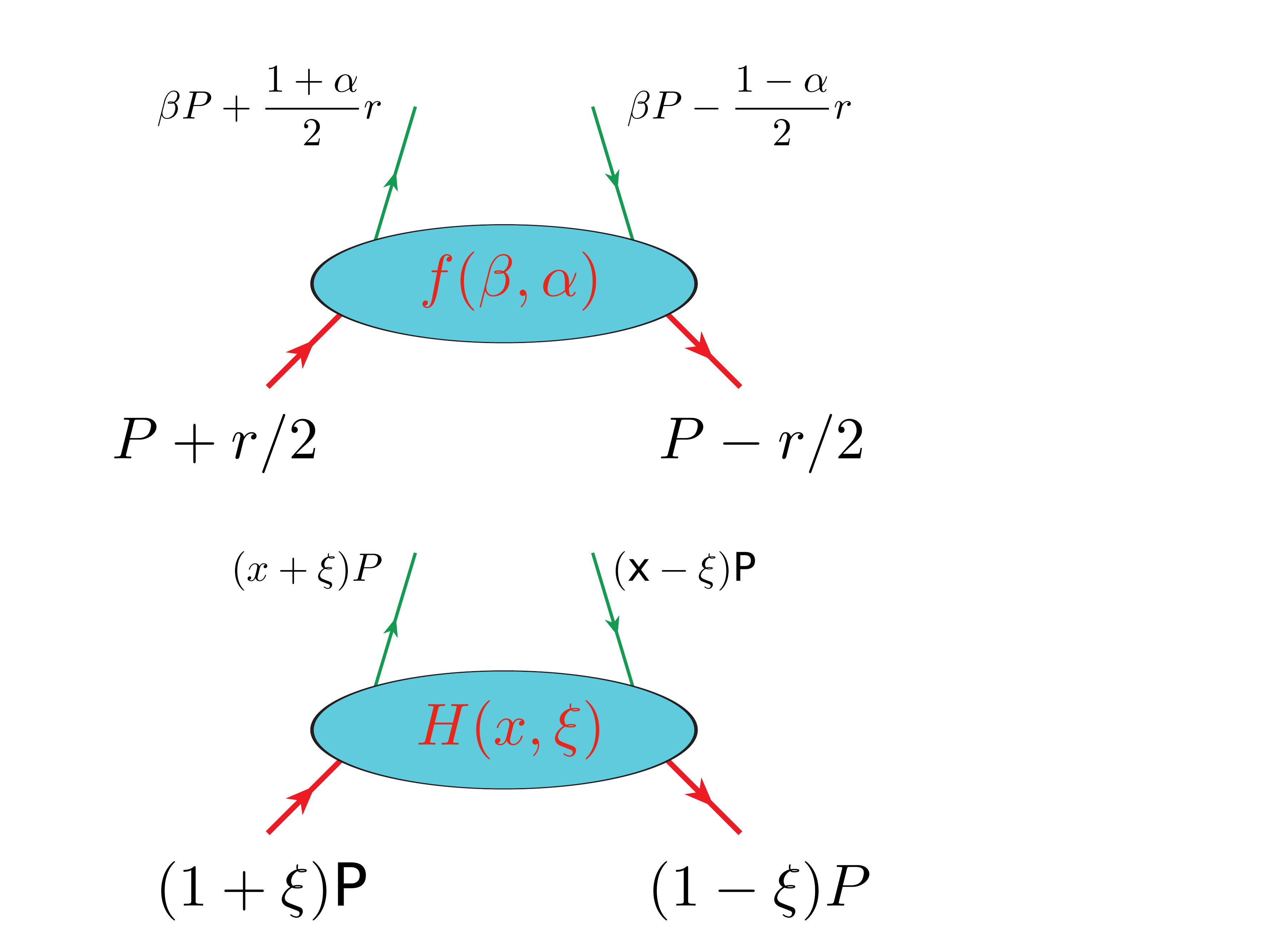} 
\end{center}
\vspace{-5mm}
\caption{Parton picture in terms of DDs and in terms of GPDs.}
\label{GPD1}
\end{figure}
\begin{align}
& H(x,\xi) =  
\int_{\Omega}  F(\beta, \alpha) \,  \delta (x - \beta -\xi \alpha)  \, d\beta \, d\alpha  \  ,
\label{GPD}
\end{align}
one can  write the handbag  contribution as 
\begin{align}
&  T(q,P,r) = 
\int_{-1}^1 H(x,\xi)  \,  D \bigl (q+(x+\xi)P \bigr )  \, dx \   .
\label{ComptGPD}
\end{align}
One may try  to  define  GPDs directly:
\begin{align}
&  \langle P+r/2 |   \psi(-z/2) \psi (z/2)|P-r/2 \rangle \nonumber \\  & =  
\int_{-1}^1 \,  e^{-ix(Pz) } H(x,\xi) \, dx +  {\cal O} (z^2) \   . 
\label{GPDdef}
\end{align}
However, an immediate question is  what is  the skewness  $\xi$  
in this definition?  It  cannot be treated as  the ratio
of $(rz)/2$  and $(Pz)$, since the ratio $(rz)/2(Pz)$  cannot be 
the same  for all points $z$. Hence, it is impossible  to 
straightforwardly use such a   definition 
in the expression (\ref{Comp})  involving a \mbox{4-dimensional} 
integration over $z$. 
But, if one uses  the DD parametrization and     integrates   over $z$, then 
the scalar products $(Pz)$  and $(rz)$  convert into the scalar products $(Pq')$
and $(rq')$, respectively,  since all other invariants, $P^2,r^2, (Pr)$ are neglected
when they appear in the ratios with $(Pq')$, $(rq')$ or $Q^2$ \footnote{DDs  and GPDs
depend on the momentum transfer \mbox{$t=r^2$,} but this dependence
is not important for our purposes.
So, in what follows, we consider the formal $t=0$ limit.}.
In this sense, only the $q'$  part of $z$ is  visible in the final  result,
and one may  define GPDs by the formula (\ref{GPDdef}) in which 
$z$ is substituted by a light-like vector $n$ proportional to $q'$, say,  by
$n_\mu = q'_\mu/2(Pq')$. 

Still,  the  appearance of  process-dependent quantities 
like $(rq')$ and $(Pq')$  in the definition of
GPDs  confronts   the basic idea  of  the  factorization
approach  that the parton distributions are
process-independent functions. 
The  standard ``escape''  is that   
 $(rq')/(Pq')$ in the GPD definitions  is  substituted by an 
 apparently  ``process-neutral''  ratio $r^{+}/P^{+}$,
supplied by information that 
$P$  basically defines the ``plus direction''   
and that  some  vector   $n$   defines the ``minus direction''
(for DVCS,   $n \sim q'$).
But this procedure creates a wrong impression that 
the definition of GPDs requires a reference to a particular frame.
As shown above, one can define GPDs $H(x,\xi)$
 through formulas (\ref{DDF}), (\ref{GPD}), which do not 
 refer  to any particular frame or process,
 and $\xi$ is just some parameter. 
 Of  course,  for each particular process, 
 $\xi$ should be adjusted to the kinematics of the process,
 e.g., $\xi=(rq')/2(Pq')$ for DVCS.
 Also, the parton interpretation of GPDs 
 has the  most natural form in the frame,
 where $P$ for a massless hadron (and $t=0$)
 defines  the plus direction.

\subsection{$D$-term}

\subsubsection{Scalar quarks}

Parameterizing the matrix element (\ref{scalarOn}), one may wish
to separate the $A_{nn} $  terms 
that are accompanied by tensors built 
from the momentum transfer vector  $r$  only
(and,  thus, invisible in the forward $r=0$ limit),
and    introduce
the {\it $D$-term} \cite{Polyakov:1999gs} 
\begin{align}
\int_{-1}^1  D(\alpha)  \,  (\alpha/2)^n \, d\alpha= A_{nn} 
\end{align}
as a function whose $(\alpha/2)^n$ moments  give $A_{nn}$.
Within the DD-parameterization,  the separation of the $D$-term
can be made   by simply using $e^{-i\beta(Pz)}= [e^{-i\beta (Pz)}-1] +1$.
The $D$-term is then given by
\begin{align}
D(\alpha) =   \int_{-1+|\alpha|}^{1-|\alpha|}  F(\beta,\alpha)\,  d \beta  \  ,
\end{align}
and the DD-parameterization converts 
into a \mbox{``DD plus D''}  parameterization
\begin{align}
&  \langle P-r/2 |   \psi(-z/2) \psi (z/2)|P+r/2 \rangle  \nonumber \\  &
=  
\int_{\Omega}  [F(\beta, \alpha)]_+ \,  e^{-i \beta (Pz) -i\alpha (rz)/2} \, d\beta \, d\alpha
 \nonumber \\  &
+  \int_{-1}^1  D(\alpha)  \,  e^{-i\alpha (rz) /2} \, d\alpha  +  {\cal O} (z^2) \   ,
\label{DDplusD}
\end{align}
where
\begin{align}
[F(\beta, \alpha)]_+  = 
 F(\beta, \alpha) -\delta (\beta)
   \int_{-1+|\alpha|}^{1-|\alpha|}
F(\gamma,\alpha)\,  d \gamma \  . 
\end{align}
is the DD with subtracted $D$-term.   Mathematically,    $[F(\beta, \alpha)]_+$ is a ``plus  distribution'' 
 with respect to $\beta$. It satisfies the condition
 \begin{align}
  \int_{-1+|\alpha|}^{1-|\alpha|} [F(\beta, \alpha) ]_+  \, d\beta = 0  \ ,
\end{align}
guaranteeing that  no $D$-term  can be constructed from $ [F(\beta, \alpha)]_+$.

\subsubsection{Spin-1/2 quarks: two-DD representation}

In the simple model with scalar quarks  discussed above, 
 one  may  just  use  the original DD  $F(\beta,\alpha)$
 without splitting it into the ``plus'' part and the $D$-term.   
In    models  with spin-1/2  quarks, it is  more difficult to avoid 
an  explicit introduction  of   extra functions producing  a $D$-term. 
The basic  reason  \cite{Polyakov:1999gs}  is  that the matrix element   of the bilocal  operator 
in that  case \footnote{Here and below we consider, for simplicity,
spin-0 hadrons.}  should  have two parts
 \begin{align}
&  \langle P-r/2 |   \bar \psi(-z/2) \gamma_\mu \psi (z/2)|P+r/2 \rangle  |_{\rm twist-2}
 \nonumber \\  &
=  2P_\mu f \bigl ((Pz),(rz),z^2 \bigr )  + r_\mu g\bigl ((Pz),(rz),z^2\bigr )   \  .
\label{pmurmu}
\end{align}
This  suggests  to introduce a   parametrization  with  two DDs
corresponding to $f$   and $g$ functions  \cite{Polyakov:1999gs}.  
For the matrix element   (\ref{pmurmu})
multiplied by $z^\mu$  -- which is  exactly what one obtains doing  the  leading-twist 
factorization  for the Compton amplitude \cite{Balitsky:1987bk} --   this gives 
 \begin{align}
& z^\mu  \langle P-r/2 |   \bar \psi(-z/2) \gamma_\mu \psi (z/2)|P+r/2 \rangle 
\nonumber 
\\  &
=  
\int_{\Omega}   e^{-i \beta (Pz) -i\alpha (rz)/2} \,\biggl [ 2(Pz)  F(\beta, \alpha) 
\nonumber 
\\  & + (rz)   G(\beta, \alpha) \biggr ]   \, d\beta \, d\alpha \ 
+{\cal O} (z^2)  .  
\label{twoDD}
\end{align}
The separation into $F$-  and $G$-parts in this  case 
is  not unique:  expanding the exponential 
in  powers of $(Pz)$ and $(rz)$, one may obtain the
same  $(Pz)^m (rz)^l$  term  both from the $F$-type and $G$-type 
parts. This leads to possibility of ``gauge transformations'' \cite{Teryaev:2001qm}:  one can 
change  
 \begin{align}
 F(\beta, \alpha) &\to F(\beta, \alpha) +\partial \chi (\beta,\alpha)/\partial \alpha  \ , 
 \label{Fgauge} \\
G(\beta, \alpha) &\to G(\beta, \alpha) -\partial \chi (\beta,\alpha)/\partial \beta  \ ,
\label{Ggauge} 
\end{align} 
using a gauge function $\chi (\beta,\alpha)$ that is odd in $\alpha$.
Still,  the terms $(Pz)^0 (rz)^l$ cannot be produced from the $F$-type  contribution.
The maximum of what   can  be done  is to absorb  all $m \neq 0$ contributions into the $F$-type term.
As a result, 
 Eq.~(\ref{twoDD}) is 
converted  into  a  ``DD  plus D''  parameterization  \cite{Polyakov:1999gs} 
in which the  term in  the square brackets is  substituted by  the 
 \begin{align} 
 2 (Pz) F_{D}(\beta, \alpha)  +  (rz) \delta (\beta) D(\alpha)
 \end{align}  
 combination, with 
$D(\alpha)$   given by the $\beta$-integral of $G(\beta,\alpha)$  and $F_{D} (\beta, \alpha)$ 
related  to the original DDs  through   the gauge transformation with 
 \begin{align} 
\chi_{D} (\beta,\alpha) = \frac12 \left \{ \int_{-\beta}^{\beta}G(\gamma,\alpha) \, d \gamma -
\int_{-1+|\alpha|}^{1-|\alpha|} G(\gamma,\alpha) \, d \gamma \right \} 
\label{D_gauge}
 \end{align}  
(cf. \cite{Teryaev:2001qm,Tiburzi:2004qr}).

\subsubsection{Spin-1/2 quarks: single-DD representation}

In fact, since the Dirac  index $\mu$  is symmetrized in the  local twist-two operators
\mbox{$\bar \psi \{\gamma_\mu  \stackrel{\leftrightarrow}{\partial}_{\mu_1} \ldots  \stackrel{\leftrightarrow}{\partial}_{\mu_n}\} \psi$}
with the $\mu_{i}$ indices related to the derivatives,
one may expect that it also produces the factor  $\beta P_{\mu}+ \alpha r_{\mu}/2$.
As shown by the authors of  Ref.~\cite{Belitsky:2000vk}, this is  precisely  what happens.
In their construction, not only the exponential produces the $z$-dependence  in the
combination $\beta (Pz) +\alpha (rz)/2$,
but also the pre-exponential terms  come  in the 
\mbox{$\beta (Pz) +\alpha (rz)/2$}  combination, i.e.,  the  result is 
a representation in which 
\begin{align}
 &2(Pz)  F(\beta, \alpha)  + (rz)   G(\beta, \alpha) \nonumber \\  & = [ 2\beta (Pz) +\alpha (rz)]
 f(\beta, \alpha)  \   , 
\end{align}
that  corresponds to $F(\beta,\alpha) =\beta f(\beta,\alpha)$
 and   \mbox{$G(\beta,\alpha) = \alpha f(\beta,\alpha)$. }  
Thus, formally,  one deals with just one DD  $f(\beta,\alpha)$. 
In principle, though,  this  single function may be a sum of  several  components, 
e.g., $\delta (\alpha) f (\beta)/\beta + \delta (\beta) D (\alpha)/\alpha$
(the result of the pioneering $D$-term paper  \cite{Polyakov:1999gs}  for
the pion DD in an effective  chiral model corresponds to $f^{I=0} (\beta,\alpha)= \delta (\alpha)/|\beta| - 
\delta(\beta)/|\alpha|$).

In the two-DD approach,  GPDs are introduced through
\begin{align}
& H(x,\xi) =  
\int_{\Omega}  \bigl [F(\beta, \alpha) +\xi G(\beta,\alpha) \bigr ] \,  \delta (x - \beta -\xi \alpha)  \, d\beta \, d\alpha  \  ,
\label{GPDFG}
\end{align}
which  converts into 
\begin{align}
H(x,\xi) &=  
\int_{\Omega}  (\beta  +\xi  \alpha)  f(\beta,\alpha) \,  \delta (x - \beta -\xi \alpha)  \, d\beta \, d\alpha
\nonumber \\
&= x \int_{\Omega}    f(\beta,\alpha) \,  \delta (x - \beta -\xi \alpha)  \, d\beta \, d\alpha  
\label{GPDf}
\end{align}
in the ``single-DD'' formulation.
The  $D$-term  in the single-DD case   is given by
\begin{align}
D(\alpha) = \alpha \int_{-1+|\alpha|} ^{1  -|\alpha|} f(\beta,\alpha) \, d\beta  \   ,
\label{DTsingle}
\end{align}
and one may  write $f(\beta,\alpha)$ as a sum 
\begin{align}
  f(\beta,\alpha)= [f(\beta,\alpha)]_+  +\delta (\beta) \frac{D(\alpha)}{\alpha}  
\end{align}
of its ``plus'' part
\begin{align}
 [f(\beta,\alpha)]_+=   f(\beta,\alpha) - \delta (\beta)  
  \int_{-1+|\alpha|} ^{1  -|\alpha|} f(\gamma,\alpha) \, d\gamma
  \label{Plussingle}
\end{align}
and $D$-term part $ \delta (\beta) {D(\alpha)}/{\alpha}$.

Despite the fact that the ``plus'' part and the $D$-term are obtained
from the same DD $f(\beta,\alpha)$,
 they are independent 
in the sense that   the 
 ``plus'' part does not contribute in Eq.~(\ref{DTsingle}),
 and 
the $D$-term  contribution
drops from Eq.~(\ref{Plussingle}).

\subsubsection{Getting GPDs from DDs}

The forward limit  $r=0$   corresponds to $\xi=0$, and GPD  $H(x,\xi)$ converts into the usual 
parton distribution $f(x)$. Using DDs, we may write
\begin{align}
f(x) &= \int _{-1+|x|} ^{1  -|x|} 
 F(x, \alpha)   \, d\alpha  = 
 x     \int_{-1+|x|} ^{1  -|x|}  f(x,\alpha) \,   d\alpha   \   .
\label{GPDtofx}
\end{align}
Thus,  the forward  distributions $f(x)$  are  obtained by integrating DDs
over vertical lines $\beta =x$  in the $(\beta,\alpha)$ plane.
For nonzero $\xi$, GPDs  are obtained from  DDs  through  integrating them 
along   the lines $\beta=x-\xi \alpha$ having   $1/\xi$ slope,
i.e. the family of $H(x,\xi)$  functions for different values of $\xi$
is obtained by ``scanning'' the same DD 
at different  angles.

\begin{figure}[h]
\begin{center} 
\includegraphics[scale=0.25]{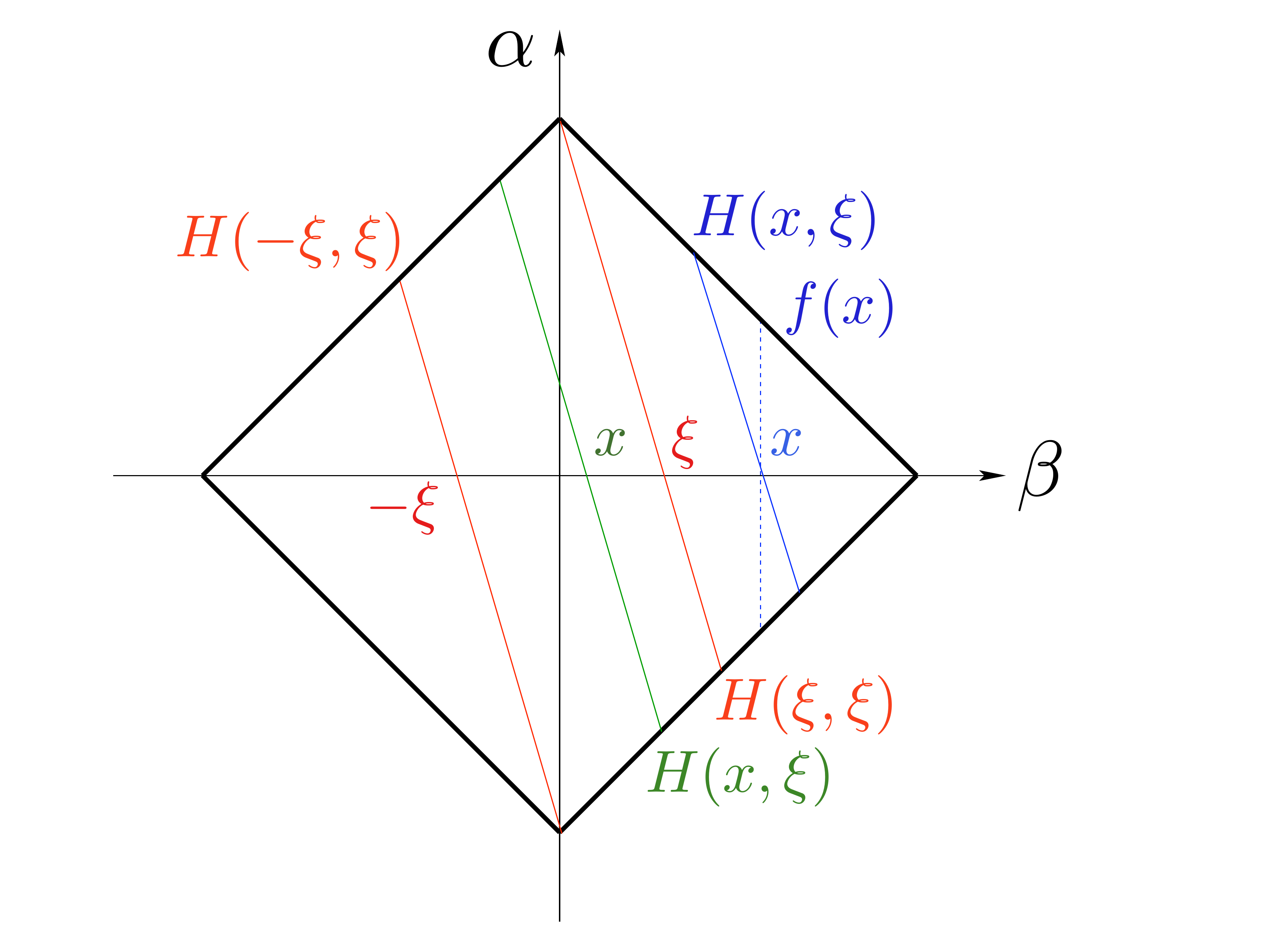} 
\end{center}
\vspace{-10mm}
\caption{Support   region for double distributions
and lines producing $f(x), H(x,\xi)$ (for $x>\xi$ and $x<\xi$),  $H(\xi,\xi)$  and $H(-\xi,\xi)$.}
\label{Support}
\end{figure}

 For $x>\xi >0$, the integration  lines  lie completely 
inside the right half of the rhombus.
The line producing GPD at  the ``border''  point  $x=\xi$  
starts  at its  upper   corner,  while  the lines corresponding to
$|x|<\xi$   cross the line $\beta =0$.  
Thus, one deals with the ``outer''  regions  $x>\xi$ and
$x<-\xi$ (in this case,   the whole  line is in the left half of the rhombus)
and the  central region $-\xi <x<\xi$,  when the integration
lines in the $(\beta, \alpha)$  plane 
lie  in  both halves of the rhombus and  intersect the $\beta =0$ line.

In GPD variables $(x,\xi)$,    the  momentum fraction  $x-\xi$ 
carried by the final  quark is positive for the right outer  region,
and negative  for the central region, i.e., in the latter case
it should be interpreted as an  outgoing antiquark
rather than incoming quark \cite{Radyushkin:1996nd},
i.e. GPD in the central region describes
emission of a quark-antiquark  pair with 
total  plus momentum $r^+$  shared in 
fractions $(1+x/\xi)/2$ and $(1-x/\xi)/2$,
like in a meson distribution amplitude.

From this physical interpretation,
one may expect that the behavior of a GPD
$H(x, \xi)$  in the central region 
is  unrelated to that in the outer region. 
But,  since  the GPD in  both regions  is obtained from the same  DD,
one  may  expect, to the  contrary, that the set of GPDs for all ``outer''  $x$'s 
and all $\xi$'s 
contains the same information as the set of GPDs  for all central
$x$'s  and all $\xi$'s.  This  ``holographic'' 
picture (cf. \cite{Kumericki:2007sa,Kumericki:2008di}) may be violated  
by terms contributing to GPDs   in the  central
region and not contributing to GPDs in the outer regions:
the terms with support on the $\beta =0$ line, i.e., those proportional
to  $\delta (\beta)$ (and, in principle,  its  derivatives), in particular, the $D$-term.
For this reason,  the usual  approach is to build separate  models 
for the $D$-term and for the remaining part of DD.  

  \subsubsection{Factorized DD Ansatz}

The reduction  formula (\ref{GPDtofx}) suggests a  model 
\begin{align}
f(\beta,\alpha) = \frac{f(\beta)}{\beta} h(\beta,\alpha)   \   ,
\label{FDDA}
\end{align}
where $f(\beta)$   is the forward  distribution, while  $h(\beta,\alpha) $  determines DD profile in the 
$\alpha$  direction  and satisfies the normalization condition
\begin{align}
 \int_{-1+|\beta |} ^{1  -|\beta|}     h(\beta,\alpha) \,   d\alpha =1\   .
\label{hnorm}
\end{align}
Since the plus component of the momentum
transfer $r$ is  shared between the quarks
in fractions $(1+\alpha)/2$ and $(1-\alpha)/2$,
like in a meson distribution amplitude,  it was proposed \cite{Radyushkin:1998es,Radyushkin:1998bz} 
to model 
the shape of the profile   function by 
\begin{align}
  h_{N}(\beta,\alpha) \sim \frac{[(1-|\beta|)^2-\alpha^2]^N}{(1-|\beta|)^{2N+1} } \ ,
  \label{hn}
\end{align}
with $N$ being a parameter governing the width of the profile.

Such a factorized DD Ansatz (FDDA) was originally  \cite{Radyushkin:1998es,Radyushkin:1998bz}
applied 
 to   an   analog of the  $F(\beta,\alpha)$ function of the two-DD
    formalism,  which  corresponds  to a model $F(\beta,\alpha) =
    f(\beta) h(\beta,\alpha)$  and $G(\beta,\alpha) =0$.  
    Later, it was corrected  by addition of the $D$-term 
    \cite{Polyakov:1999gs},
    which formally corresponds to the  ``gauge''  (\ref{D_gauge}) in which 
     $G(\beta,\alpha) \to G_{D}(\beta, \alpha)=\delta (\beta) D(\alpha)$, and $F(\beta,\alpha)\to F_{D}(\beta, \alpha)$.
   Note that if  $F=\beta f$ and \mbox{$G=\alpha f$, }
     the model $F_{D}(\beta,\alpha) = f(\beta)\, h(\beta,\alpha)$ 
   does not coincide with the model $f(\beta,\alpha) = f(\beta)\, h(\beta,\alpha)/\beta$,
   since the gauge function $\chi_{D} (\beta, \alpha)$ (see Eq.~(\ref{Fgauge})) 
   is  nontrivial.

Thus, there is a question whether the FDDA should be applied 
to $F_{D}(\beta,\alpha)$ (as it was done so far)
or to the DD $f(\beta,\alpha)$ of the
single-DD formulation.  
It should  be confessed that no  enthusiasm has been observed 
 to use FDDA in the form of 
 the single-DD 
   formula (\ref{FDDA}). This observation   has a simple explanation:  the function 
    $f(\beta)/\beta$ is not integrable for $\beta =0$,
    even if $f(\beta)$ is finite for $\beta=0$.
    The reason is that the DVCS amplitude contains singlet  
    GPDs, which are odd   functions of $\beta$.
    Hence,  $f(\beta) /\beta$ should be an even  function, and 
    the principal value prescription does not work.
 Moreover, for small $\beta$ one  would expect that the
 forward distribution $f(\beta) $ has a 
 singular $f(\beta)  \sim 1/\beta^a $
 Regge  behavior, which makes the problem even worse.

\section{GPD  model with implanted Regge behavior}

\subsection{Formulation}

The assumptions  used in the factorized DD Ansatz
are based on the experience  with calculating 
DDs for  triangle diagrams \cite{Radyushkin:1998rt} 
and form factors in  the  light-front formalism models 
with power-law dependence of the wave function 
on transverse momentum \cite{Mukherjee:2002gb}
(see also \cite{Hwang:2007tb}).

\begin{figure}[h]
\begin{center} 
\includegraphics[scale=0.2]{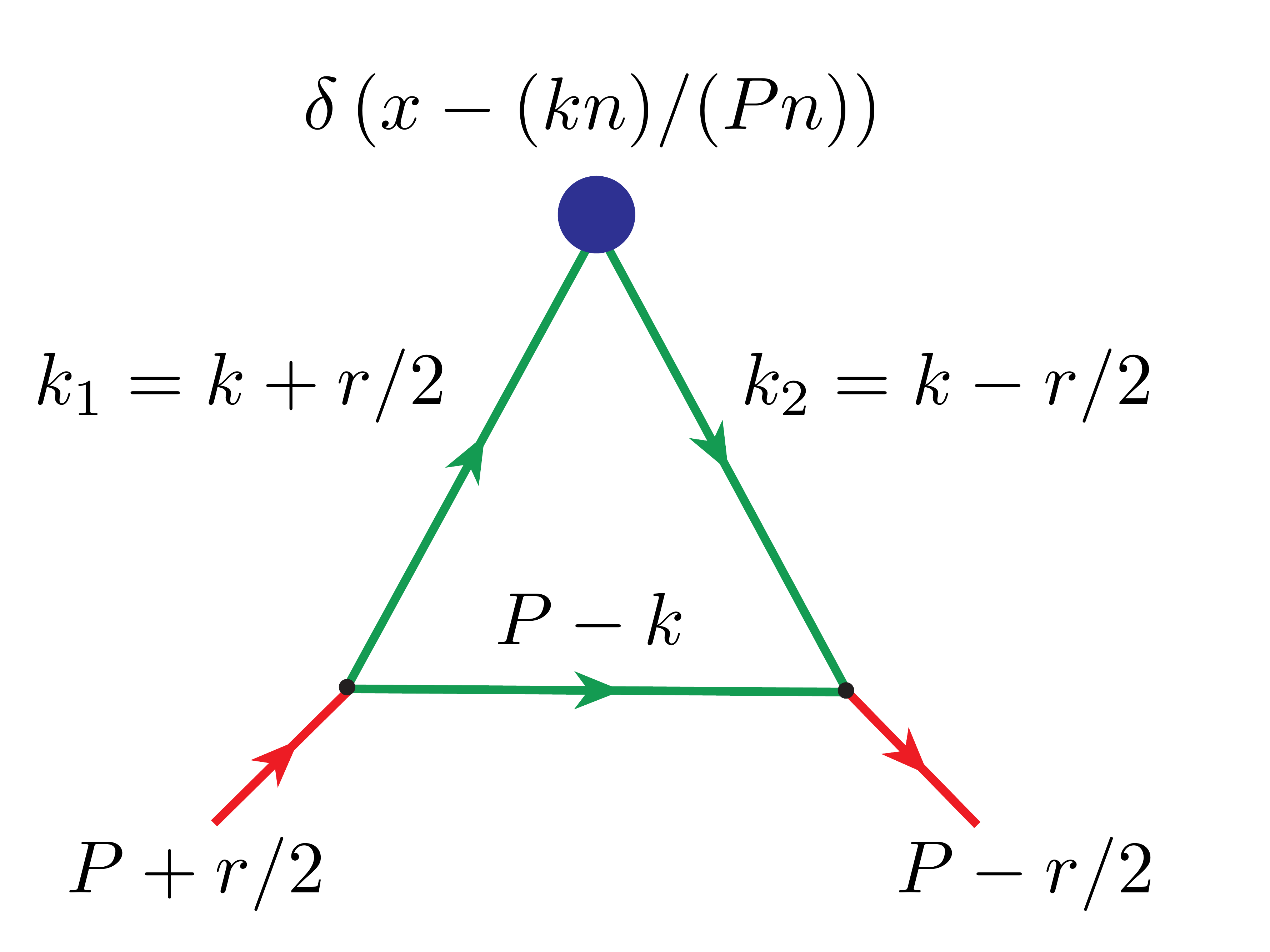} 
\end{center}
\vspace{-5mm}
\caption{Triangle  diagram  model for GPD.}
\label{triangle}
\end{figure}

The  simplest triangle  diagram  (see Fig.\ref{triangle})  in the scalar  model corresponding
to Eq.~(\ref{scalarOn})  may  be  used 
as an  example of a model for  GPD
\begin{align}
 H(x,\xi) \sim 
\int 
\frac{ d^4 k\, \delta (x- (kn)/(Pn))}{(m_1^2-k_{1}^2 )
 (m_2^2-k_{2}^2)(m_3^2-(P-k)^2 )}  \  .
\end{align}
Though the  \mbox{$\xi$-dependence} is not immediately  visible here, it appears 
after integration over $k$  through  the $(rn)/2(Pn)$  ratio.
The DD $F(\beta,\alpha)$  generated by this  diagram  is  just a constant  
\cite{Radyushkin:1997ki},
which  corresponds to a flat  $N=0$  profile \mbox{$h_0 (\beta, \alpha) \sim 1/(1- \beta)$} and
$f(\beta) \sim 1- \beta$  forward  distribution.  

The calculation \cite{Mukherjee:2002gb} of overlap integrals for light-front 
wave functions with a power-law behavior
$\psi (x, k_{\perp}) \sim  1/(k_{\perp}^2)^{1+\kappa}$ 
resulted in expressions equivalent to using DDs 
with $N=\kappa$  profile  in Eq.~(\ref{hn})
and forward distributions behaving like $(1-\beta)^{2\kappa+1}$. 
The same profile arises   \cite{Mukherjee:2002gb}  if one differentiates  
a scalar triangle diagram  $\kappa $   times with respect to
masses (squared) of each active quark.

\begin{figure}[t]
\begin{center} 
\includegraphics[scale=0.15]{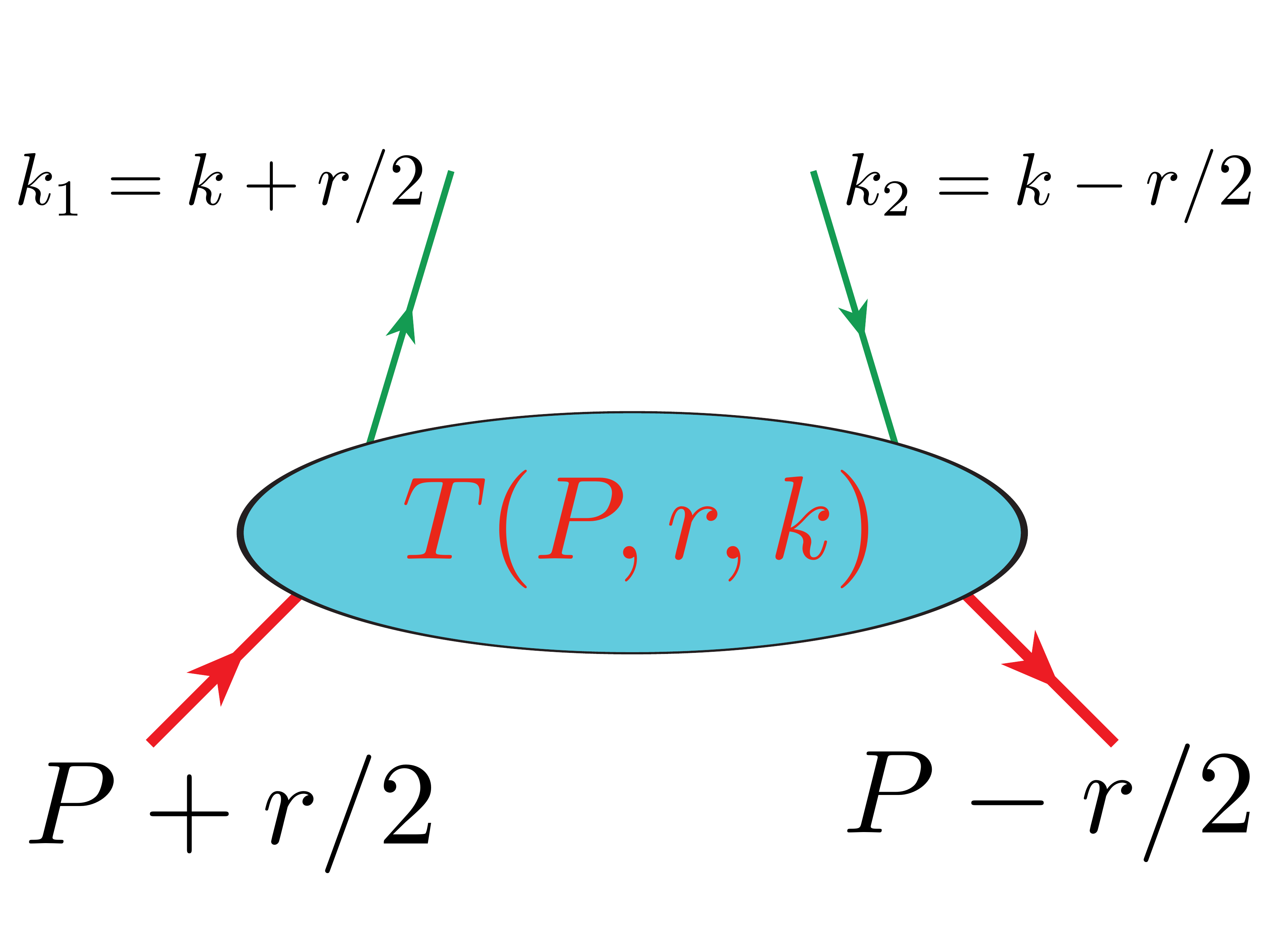} 
\end{center}
\vspace{-5mm}
\caption{Hadron-quark scattering amplitude.}
\label{blobPk}
\end{figure}

The triangle diagrams, however, do not generate 
the Regge $f(\beta) \sim 1/\beta^a$  behavior 
for small  $\beta$. The latter may be obtained, in particular,  
by infinite summation of 
 higher-order $t$-channel ladder diagrams 
 (see, e.g., \cite{Efremov:2009dx}). 
A simpler way was proposed in Ref.~\cite{Szczepaniak:2007af}, where 
 the spectator propagator     
 was substituted  by a 
parton-hadron scattering amplitude $T(P,r,k)$  (see Fig.\ref{blobPk}) written 
in the dispersion relation representation. 
To  avoid  divergencies generated by  the Regge behavior, the subtracted
dispersion relation
\begin{align}
  T(P,r,k)  \to & T((P-k)^2)      =   T_0   \nonumber \\ &  
  +
\int_0^\infty d \sigma {\rho(\sigma) \, } \left \{ \frac{1}{ \sigma -(P-k)^2}
-\frac{1}{ \sigma} \right \}
\label{DRel}
      \end{align}
was  used. The  spectral function $\rho (\sigma)$  here 
should be adjusted to  produce a 
  desired  Regge-type behavior with respect to $s=(P-k)^2$
  \footnote{To get the $s^{a}$ Regge behavior with $1<a<2$,
  one should use a doubly subtracted dispersion relation,
  but in this paper we will follow 
  the original construction of Ref.~\cite{Szczepaniak:2007af},
  leaving the generalization for $a>1$ to a future work.}.

In the light-front formalism, the starting contribution
corresponds to a triangle diagram in which
the hadron-quark vertices are substituted
by the light-front wave functions $\psi (x, k_{\perp})$
that  bring in an extra fall-off of the integrand at large transverse 
momenta $k_{\perp}$. 
The authors of Ref.~\cite{Szczepaniak:2007af} intended 
to reflect this physics  in their 
covariant model.
To introduce form  factors bringing 
in a faster fall-off   of the $k$-integrand with respect to
  quark  virtualities 
$k_1^2$ and $k_2^2$,   it was  proposed to 
use higher powers of $1/(m_i^2-k_i^2)$  
instead  of perturbative propagators,
which may be achieved by differentiating
the triangle diagram with respect to $m_i^2$.

The model of Ref.~\cite{Szczepaniak:2007af}
assumes spin-1/2 quarks.
It was  argued that the  Dirac structure of  the 
hadron-parton scattering amplitude
in  this case  should be   given by   $\slashed k$,
which 
provides  EM gauge  invariance of  the  DVCS  amplitude. 
Summarizing,  the model   scattering amplitude  
has the following structure 
\begin{align}
  \, \frac{ \slashed k \  T((P-k)^2}{(m_1^2-k_1^2)^{N_1+1}(m_2^2-k_2^2)^{N_2+1}}  \ .
\label{QHSA}
\end{align}
To   treat the two  quarks on   equal footing, it makes sense to take
$m_1=m_2$,  but we will keep them  different for a  while,
to  separate effects produced by nonzero $N_1$  and $N_2$.

After $\slashed k$ is  contracted with the $\slashed n$  factor from  the operator vertex,
one  gets $(kn)$,  and the model GPD  that will be analyzed below is given  by 
\begin{widetext}
\begin{align}
 H(x,\xi) = \frac1{\pi^2} \frac{N_1! N_2!}
{(N_1+N_2)!}
 &
\int \frac{ (kn) }{(Pn)}
\frac{ d^4 k\, \delta (x- (kn)/(Pn))}{[m_1^2-(k+r)^2 ]^{N_1+1}
 [m_2^2-(k-r)^2]^{N_2+1}}
 \left [T_0 +
\int_0^\infty d \sigma {\rho(\sigma) \, } \left \{ \frac{1}{\sigma-(P-k)^2 }
-\frac{1}{ \sigma} \right \} \right ] \  .
\end{align}
The overall factors were introduced here  for  future convenience.
Using the $\alpha$-representation 
\begin{align}
 \frac{N!}{(m^2-k^2)^{N+1}}
=  \int_0^{i \infty} e^{\alpha  (k^2-m^2)}
\alpha^N \, d\alpha
\hspace{1cm} , \hspace{1cm}
\frac{1}{\sigma}
=  \int_0^{i \infty} e^{-\alpha_3 \sigma}
 \, d\alpha_3
\end{align} 
 for  propagators  and also for the $1/\sigma$ subtraction term
gives
\begin{align}
\frac{x}{(N_1+N_2)!}   &  \int_0^{\infty}  d \sigma \,  \rho(\sigma)   \int_0^{i \infty}
{\alpha_1^{N_1} d\alpha_1 \, 
\alpha_2^{N_2} d\alpha_2 \,   d\alpha_3} \, 
  \left \{ \delta \left (x- \frac{ \alpha_3+(\alpha_2-\alpha_1)(rn)/2(Pn)}
{\alpha_1+\alpha_2+\alpha_3} \right )  \frac1{(\alpha_1+\alpha_2+\alpha_3)^2} \right. \nonumber \\ &  \left.  -
 \delta \left (x- \frac{(\alpha_2-\alpha_1)(rn)/(Pn)}
{\alpha_1+\alpha_2} \right ) \, 
\frac{1}{(\alpha_1+\alpha_2)^2} \right \}e^{-\alpha_3  \sigma-\alpha_1 m_1^2 -  \alpha_2 m_2^2 } 
\end{align}
for the terms involved in the dispersion integral. 
The second delta-function corresponds to the $1/\sigma$ subtraction term
of the dispersion representation. 
It is accompanied by the $1/(\alpha_1+\alpha_2)^2$  factor   because 
$1/\sigma$ does not have $k$-dependence.
 Introducing  the skewness variable $\xi \equiv (rn)/2(Pn)$, changing $\alpha_i \equiv x_i \lambda$ and integrating over $\lambda$ we obtain 
\begin{align}
  x    \int_0^\infty d \sigma \,  \rho(\sigma)  &  \int_0^1 
\frac{x_1^{N_1} dx_1 \, 
x_2^{N_2} dx_2 \,   dx_3 \, \delta (1- x_1-x_2-x_3)}{ 
(x_3  \sigma+x_1 m_1^2+x_2 m_2^2)^{N_1+N_2+1} } 
 \left \{  \delta \left (x- x_3-{(x_2-x_1)\xi} \right ) 
- \frac{\delta \left (x- {(x_2-x_1)\xi} \right )}{(x_1+x_2)^2}  
 \right \}  \  .
\label{eq5} 
\end{align}
Taking equal masses $m_1=m_2 \equiv m$, using $x_1+x_2=(1-x_3)$  and introducing $z$
through $x_1=(1-x_3)z$ 
results in 
\begin{align}
  x    \int_0^\infty d \sigma \,  \rho(\sigma)  &  \int_0^1 dx_3 \, dz \, 
\frac{(1-x_3)^{N_1+N_2+1} 
z^{N_1} (1-z)^{N_2} \,}{ 
[x_3  \sigma+(1-x_3) m^2]^{N_1+N_2+1} }  
 \left \{  \delta \bigl (x- x_3-{(x_2-x_1)\xi} \bigr ) 
- \frac{\delta \left (x- {(x_2-x_1)\xi} \right )}{(1-x_3)^2}  
 \right \}  \   .
\label{eq7} 
\end{align}
The $T_0$  subtraction  term   gives the  $D$-term-type  contribution
\begin{align}
  D_0 (x/\xi) =    \frac{T_0}{2^{N_1+N_2} (N_1+N_2)}  \left (  \frac{x}{|\xi|} 
\right )\left ( 1- \frac{x}{\xi} \right )^{N_1} 
\left ( 1+ \frac{x}{\xi} \right )^{N_2} \, \theta \left ( \left | \frac{x}{\xi}\right | <1 \right ) 
\label{eq7T0} 
\end{align}
that  vanishes outside the central region and,  hence,  is
invisible in the forward limit.  In  what  follows,   we will concentrate on the terms generated 
by the dispersion integral,  but 
one should   remember that the $D_0$   term can  always be added
to GPD $H(x,\xi)$, i.e.,  in all formulas below one
should be ready to change $H(x,\xi) \to H(x,\xi)  +  D_0 (x/\xi)$.

\subsection{Forward  case}

 The case $\xi =0$ corresponds to the   forward distribution
\begin{align}
 H(x, \xi=0) =  x    \int_0^\infty d \sigma \,  \rho(\sigma)  &  \int_0^1 dx_3 \, dz \, 
\frac{(1-x_3)^{N_1+N_2+1} 
z^{N_1} (1-z)^{N_2} \,}{ 
[x_3  \sigma+(1-x_3) m^2]^{N_1+N_2+1} } 
\left  \{  \delta  (x- x_3)  
- \frac{\delta (x )}{(1-x_3)^2}  
\right \}  \  .
\label{eq8} 
\end{align}
Taking   $x \delta (x) =  0$  for $x \neq 0$  gives 
\begin{align}
 H(x, \xi=0)  = & \frac{N_1!N_2!}{(N_1+N_2+1)!}  \, x \,  (1-x)^{N_1+N_2+1} \int_0^\infty   
\frac{d \sigma \,  \rho(\sigma) }{ 
(x  \sigma+ (1-x)    m^2)^{N_1+N_2+1} } 
 \equiv f(x)  \  .
\label{eq_forw}
\end{align}
Formally, we may  also write
\begin{align}
 \frac{H(x, \xi=0)}{x} = \frac{ f(x)}{x}   
- {\delta (x )} \int_0^1 dx_3 \,  \frac{f(x_3)}{x_3(1-x_3)^2} 
\label{eq81} 
\end{align}
The second term provides the subtraction   regularizing   the function 
$f(x)/x$ at  its 
 singular point $x=0$.

\subsection{DD description} 

In the double distribution representation,  we have 
$x=\beta + \alpha \xi$.
So, turning back to Eq. (\ref{eq5}) and changing there
$1-x_1-x_2 \equiv \beta$, $x_2-x_1 \equiv \alpha$,  we obtain  that 
\begin{align}
 x_1 = \frac12 (1-\beta -\alpha) 
\   \  , \  \  
 x_2 = \frac12 (1-\beta +\alpha)  \ ,
\end{align}
which gives for equal masses 
\begin{align}
  \frac{x}{2^{N_1+N_2+1} }  \int_0^\infty d \sigma \,  \rho(\sigma)  
&  \int_0^1 d\beta \int_{-1+\beta}^{1-\beta} d \alpha \, 
\frac{(1-\beta -\alpha)^{N_1} \, 
(1-\beta +\alpha)^{N_2} }{ 
(\beta  \sigma+(1-\beta) m^2)^{N_1+N_2+1} } 
 \left \{  \delta \left (x- \beta- \alpha \xi  \right ) 
- \frac{\delta \left (x- \alpha \xi  \right )}{(1-\beta)^2}  
 \right \}  \  .
\label{eq_alpha} 
\end{align}
Thus,  a  faster  decrease of the $k$-integrand 
with respect to  the quark  virtualities 
$k_1^2$ or $k_2^2$   results in a  suppression 
of the DD behavior by  powers  of $(1-\beta -\alpha)$ or 
$(1-\beta +\alpha)$ when $\alpha$  approaches 
the support boundary $|\alpha|=1-\beta$.
For equal $N_1=N_2=N$, we  obtain  
\begin{align}
  \frac{x}{2^{2N+1} }  \int_0^\infty d \sigma \,  \rho(\sigma)  
&  \int_0^1 d\beta \int_{-1+\beta}^{1-\beta} d \alpha \, 
\frac{[(1-\beta)^2 -\alpha^2]^{N} \, 
 }{ 
(\beta  \sigma+(1-\beta) m^2)^{2N +1} }  
 \left \{  \delta \left (x- \beta- \alpha \xi  \right ) 
- \frac{\delta \left (x- \alpha \xi  \right )}{(1-\beta)^2}  
 \right \}  \ .
\label{eq_nn} 
\end{align}
Using Eq.(\ref{eq_forw}), one  can substitute the $\sigma$-integral through
forward distribution to get 
\begin{align}
H(x, \xi) =  \frac{x}{2^{2N+1} }  \frac{(2N+1)!}{(N!)^2} 
 &  \int_0^1 d\beta \, \int_{-1+\beta}^{1-\beta} d \alpha \,  
\frac{[(1-\beta)^2 -\alpha^2]^{N} \, 
 }{ 
(1-\beta)^{2N +1} }\,  \frac{f(\beta)}{\beta} 
 \left \{  \delta \left (x- \beta- \alpha\xi \right ) 
- \frac{\delta \left (x- \alpha \xi \right )}{(1-\beta)^2}  
 \right \}  \ .
\label{eq_nnn} 
\end{align}
This  trick  allows one to by-pass the question 
about  the  specific form of the spectral density $\rho(\sigma)$.
 
It is  easy  to notice that the factor
\begin{align}
h_N(\beta,\alpha) \equiv   \frac{1}{2^{2N+1} }  \frac{(2N+1)!}{(N!)^2}  
\frac{[(1-\beta)^2 -\alpha^2]^{N} \, 
 }{ 
(1-\beta)^{2N +1} }
\label{eq_profile} 
\end{align}
is  precisely a normalized profile satisfying 
\begin{align}
\int_{-1+\beta}^{1-\beta} h_N(\beta,\alpha)\, d \alpha=1 \  .
\end{align}
Since $x_1+x_2 \leq 1$  for  the Feynman parameters $x_1,x_2$,
we have $\beta \geq 0$ in the expressions  above.
The $\beta \leq 0$  part of DD  comes from the crossed  diagram,
in which the dispersion relation is written for $T((P+k)^2)$.
For the singlet case, the full DD $f(\beta,\alpha)$  should  be   symmetric
with respect to interchange $\beta \to - \beta$ (and also symmetric 
under $\alpha \to - \alpha$), which results in GPD $H(x,\xi)$ that is an odd function of $x$.
For this reason, we will proceed with  the $\beta \geq 0$  case, keeping in mind  to 
antisymmetrize the resulting $H(x,\xi)$  at the very end.

Thus, we can rewrite   Eq.(\ref{eq_nnn})  as
\begin{align}
\frac{H  (x, \xi)}{x} =  &   \int_0^1 d\beta \, \int_{-1+\beta}^{1-\beta} d \alpha \,  
  \frac{f(\beta)}{\beta} \, h_N(\beta,\alpha)
\,   \left \{  \delta \left (x- \beta- \alpha\xi \right ) 
- \frac{\delta \left (x- \alpha \xi \right )}{(1-\beta)^2}  
 \right \}  \  . 
\label{eq_DDA} 
\end{align}
The first term  
\begin{align}
\frac{H^{(1)} (x, \xi)}{x} =     \int_0^1 d\beta \, \int_{-1+\beta}^{1-\beta} d \alpha \,  
  \frac{f(\beta)}{\beta} \, h_N(\beta,\alpha)
\,  \delta \left (x- \beta- \alpha\xi \right ) 
\label{eq_DDA1} 
\end{align}
coincides with the  factorized  DD  Ansatz for $H(x,\xi)/x$ in which it is  reconstructed 
from its  forward limit $f(x)/x$.
The relevant double  distribution is given by 
$
 f(\beta,\alpha) =   h_N(\beta,\alpha) {f(\beta)}/{\beta} $.
The second term may be rewritten as 
\begin{align}
\frac{H^{(2)} (x, \xi)}{x} =  -   \int_0^1 d\beta \, \int_{-1+\beta}^{1-\beta} d \alpha \,  
 \delta \left (x- \beta- \alpha\xi \right )  \, \delta (\beta)  
\int_0^{1-|\alpha|}  d \gamma \,  \frac{f (\gamma, \alpha)}
{(1-\gamma)^2} \, , 
\label{eq_DDAreg} 
\end{align}
\end{widetext}
and it provides a regularization of the $\beta$-integral in Eq. (\ref{eq_nnn}).
The total contribution is given by 
\begin{align}
\frac{H (x, \xi)}{x}& =      \int_0^1 d\beta \, \int_{-1+\beta}^{1-\beta} d \alpha \,  
 \delta \left (x- \beta- \alpha\xi \right )  \nonumber \\   \times &
\left \{ f(\beta,\alpha) -  \delta (\beta)  
\int_0^{1-|\alpha|}  d \gamma \,  \frac{f (\gamma, \alpha)}
{(1-\gamma)^2}   \right \} \  . 
\label{eq_DDAtotal} 
\end{align}

Thus, the model of Ref.~\cite{Szczepaniak:2007af}, first,
corresponds to the single-DD representation  (\ref{GPDf}), 
and, second, it has the structure 
of the factorized DD Ansatz  (\ref{FDDA}).
Furthermore, due to the subtraction in the
dispersion relation   (\ref{DRel}), one deals  with the regularized double  distribution
\begin{align}
f^{\rm  reg} (\beta, \alpha)  =   f(\beta,\alpha) -  \delta (\beta)  
\int_0^{1-|\alpha|}  d \gamma \,  \frac{f (\gamma, \alpha)}
{(1-\gamma)^2} \  .
\label{eq_DDAreg1} 
\end{align}

\begin{figure}[h]
\begin{center} 
\includegraphics[scale=0.25]{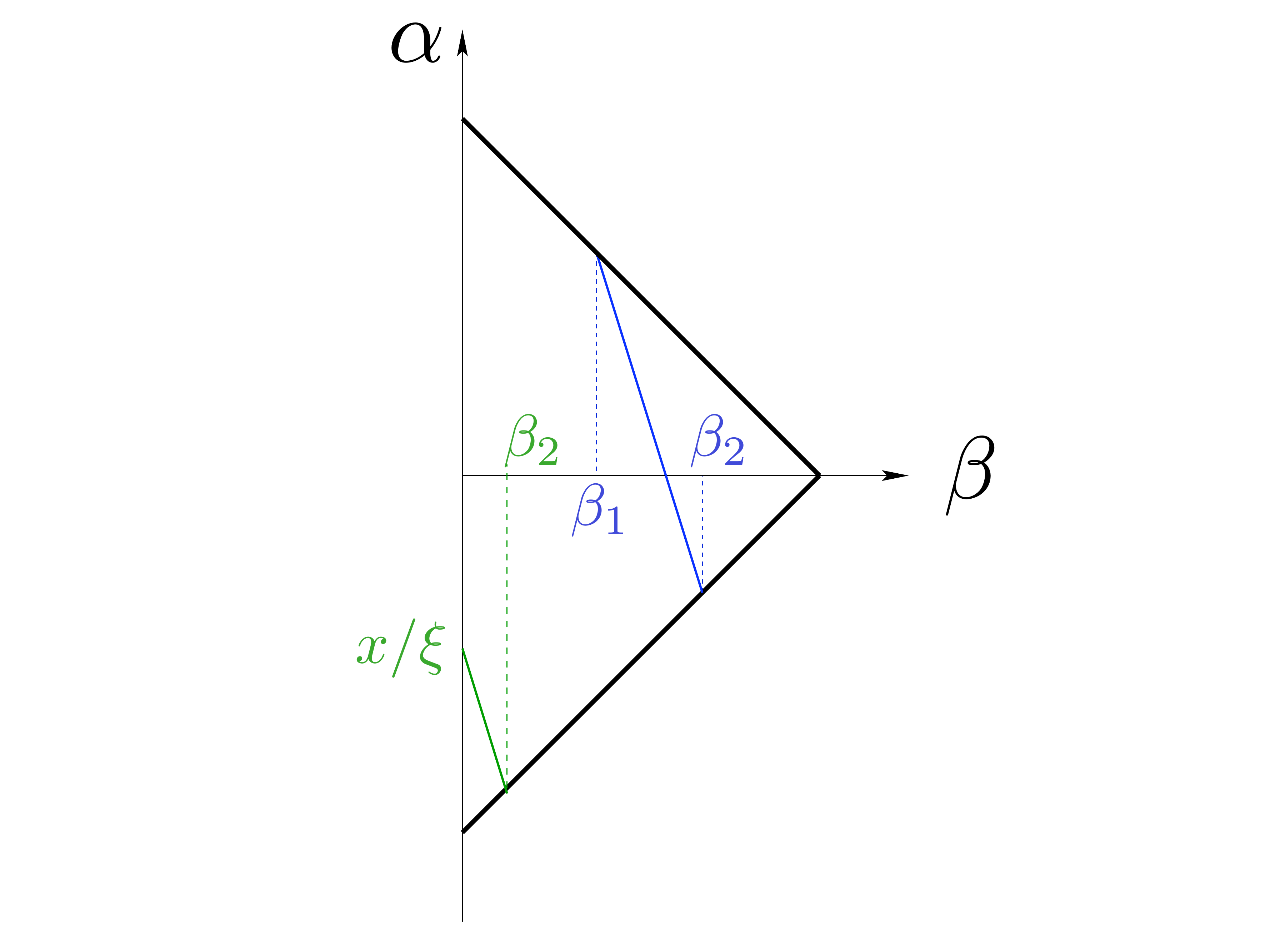} 
\end{center}
\vspace{-10mm}
\caption{Support region and integration lines  producing $H(x,\xi)$ for $x>\xi$ and $x<\xi$
from a double distribution  that is  nonzero for $\beta>0$ only.
}
\label{GPD2}
\end{figure}

Returning back to Eq.(\ref{eq_DDA})
and calculating integral over $\alpha$, we formally obtain
\begin{align}
&H  (x, \xi) =  \frac{x}{\xi} \, \theta (x>\xi)     \int_{\beta_1}^{\beta_2}  d\beta \,
  \frac{f(\beta)}{\beta} \, h(\beta,(x-\beta)/\xi)  \nonumber \\ 
&+ \frac{x}{\xi}\,  \theta (|x|<\xi)     \int_{0}^{\beta_2}  d\beta \,
  \frac{f(\beta)}{\beta} \, h(\beta,(x-\beta)/\xi) \nonumber \\ 
&- \frac{x}{\xi}\,  \theta (|x|<\xi)     \int_{0}^{1-|x|/\xi}  d\beta \,
  \frac{f(\beta)}{\beta (1-\beta)^2} \, h(\beta,x/\xi)  \  ,
\label{eq_DDA2} 
\end{align}
where (see Fig.~\ref{GPD2})
\begin{align}
 &\beta_1 = \frac{x-\xi}{1-\xi}  \ , 
 \beta_2  = \frac{x+\xi}{1+\xi}  \  .
\end{align}

 One  should  realize, however, that the second 
and  the  third integrals diverge and 
should be combined together to regularize their  singularity at $\beta=0$.   
This may be achieved by rewriting the $|x|<\xi$ part in the form 
\begin{align}
&H  (x, \xi)  |_{|x<\xi} =  
 \frac{x}{\xi}\,      \int_{0}^{\beta_2}  d\beta \,
  \frac{f(\beta)}{\beta} \, \bigl [h(\beta,(x-\beta)/\xi) -h(\beta,x/\xi) \bigr ] \nonumber \\ 
 & +  \frac{x}{\xi}\,      \int_{0}^{\beta_2}  d\beta \,
  \frac{f(\beta)}{\beta} \, h(\beta,x/\xi) \left (1- \frac1{(1-\beta)^2} \right )   \nonumber \\ 
&- \frac{x}{\xi}\,     \int_{\beta_2}^{1-|x|/\xi}  d\beta \,
  \frac{f(\beta)}{\beta (1-\beta)^2} \, h(\beta,x/\xi)
\label{eq_x<xi1} 
\end{align}
explicitly showing the compensation of the $1/\beta$ 
factor.

\subsection{Results} 

\subsubsection{$N=1$ profile}

For the model forward distribution 
\begin{align}
 f_a (\beta)= {(1-\beta)^3}/{\beta^a}
\end{align}
and the profile function
\begin{align}
 h_1 (\beta, \alpha) = \frac34 \, \frac{(1-\beta)^2 - \alpha^2}{(1- \beta)^3}  \  ,
\end{align}
we obtain, for $x>\xi$:
\begin{align}
H  (x, \xi)|_{x>\xi}  = &\frac34\,  \frac{x}{\xi}    \int_{\beta_1}^{\beta_2} 
  \frac{d\beta}{\beta^{a+1}} 
\left \{ (1-\beta)^2 - \left (\frac{x-\beta}{\xi} \right )^2 \right \} \  .
\label{eq_x>xi} 
\end{align}
Calculating $H(\xi,\xi)$, i.e.,  the GPD at the border point $x=\xi$,    
one  gets here the 
$[(1-\beta)^2-(1-\beta/\xi)^2] \sim \beta$
factor
from the profile function,
and this factor   changes the strength of singularity for $\beta =0$.
As a result, the 
integral over $\beta$   converges as  far as  $a<1$.
This outcome 
 is  a  consequence of using  a profile  function  that linearly vanishes 
at the sides of  the support rhombus. In  its turn, the
$N=1$ profile 
  is   generated by the 
assumed $1/(k_1^2 k_2^2)^2$  dependence
of the $k$-integrand  for large parton
virtualities. 
If one takes  the $N=0$ profile,    the factor in the curly 
brackets should  be substituted by $1/(1-\beta)$),    
and the integral producing $H(\xi,\xi)$ diverges.
For small, but  nonzero $x-\xi$,  one obtains the behavior   
  proportional to $1/\beta_1^a \sim (x-\xi)^{-a}$.
This   result is  similar to that obtained in Ref.~\cite{Szczepaniak:2007af}.
 However,   since its authors 
explicitly declared that they are going to  soften the hadron-quark   vertices 
by  differentiating the  diagram over  the quark  masses,
one may wonder,  how  did  it happen that they obtained a singular result?

\begin{figure}[h]
\begin{center} 
\includegraphics[scale=0.16]{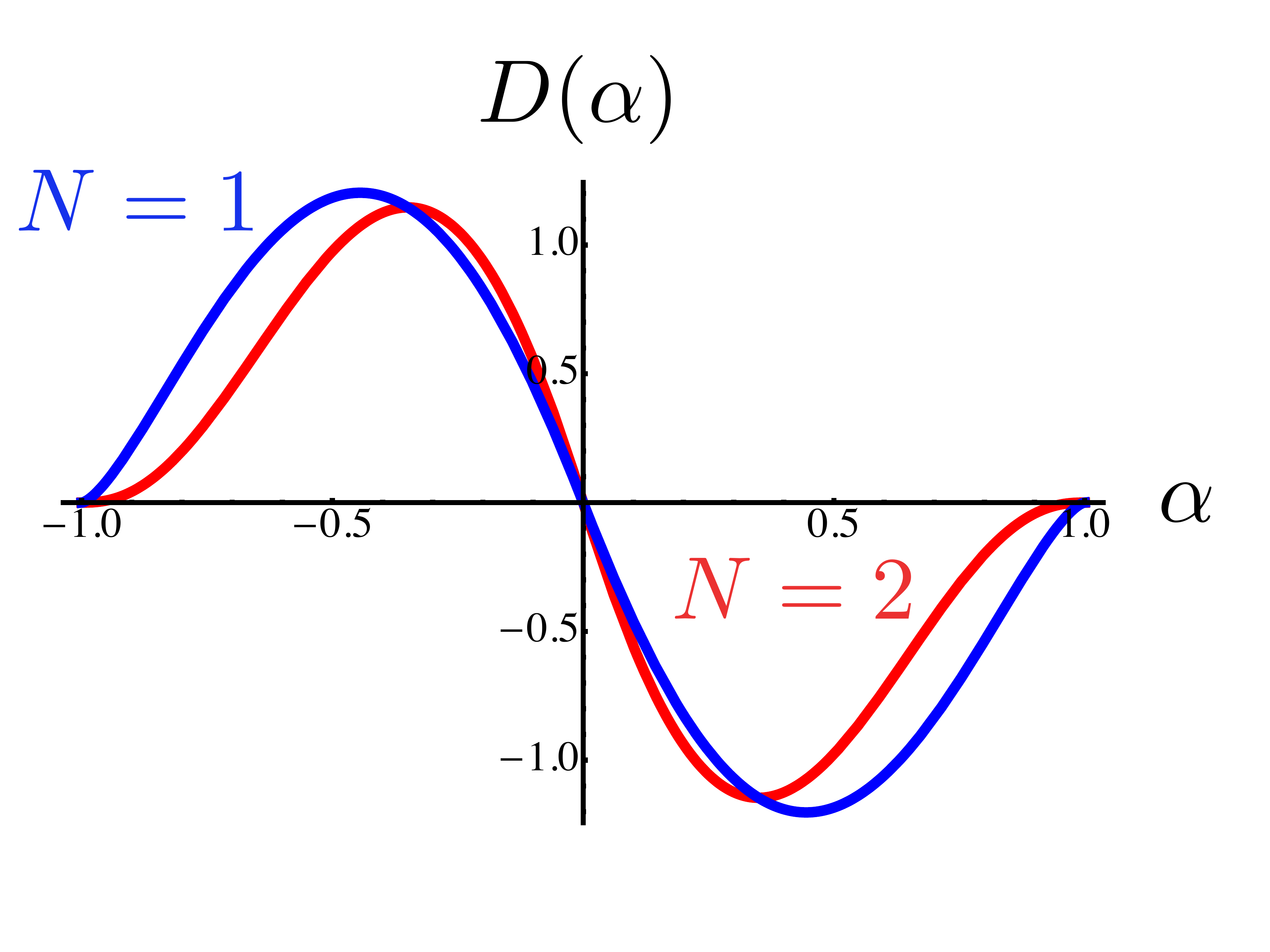} 
\vspace{-1cm}
\caption{The  $D$-terms  in  $N=1$   and $N=2$  profile models  for $a=0.5$.}
\label{dt1and2}
\end{center}
\end{figure}

The subtlety is that they  took equal quark  masses $m_1=m_2=m$
from the very beginning,  and used differentiation  with respect to this  common $m^2$.
Here it  should  be  noted that because 
\begin{widetext}
\begin{align}
\left (    \frac{d}{dm^2} \right )^2 
&
\frac1 {( m^2-k_1^2) ( m^2-k_2^2)}
= \frac1{( m^2-k_1^2)^3 ( m^2-k_2^2)  }
+ \frac2{( m^2-k_1^2)^2 ( m^2-k_2^2)^2  }
+ \frac1{( m^2-k_1^2) ( m^2-k_2^2)^3  }  \  ,
\end{align}
the first and the third  term on the r.h.s. are not  softened with respect
to one of the virtualities,  i.e., one of the hadron-parton vertices
remains pointlike. As we have seen above, imposing the $1/(k_1^2)^{N_1+1}  (k_2^2)^{N_2+1}$
dependence on   virtualities  one would obtain the
 $(1-\beta-\alpha)^{N_1}(1-\beta+\alpha)^{N_2}/2^{N_1+N_2}$
factor, i.e.,  every differentiation with respect to $m_1^2$ gives $(1-\beta-\alpha)/2$,
while every differentiation with respect to $m_2^2$ gives the $(1-\beta+\alpha)/2$ factor, both  
resulting in a  nontrivial profile in the $\alpha$-direction.
On the  other hand, each  differentiation with respect to the common $m^2$  
gives the $(1-\beta-\alpha)/2+(1-\beta+\alpha)/2=(1-\beta)$  factor that has  no dependence
on $\alpha$. This kind  of  softening  only increases the power
of  $(1-\beta)$,   but  DD remains flat in the $\alpha$-direction. 

Note, that the use of $1/(k_1^2)^{N+1}  (k_2^2)^{N+1}$-dependence 
in the model $D_{0}$-term contribution (\ref{eq7T0})  results in the \mbox{$(1-x^{2}/\xi^{2})^{N}$} 
factor, which gives $D_{0}(1)=0$ for $N>0$ case. 
This  vanishing of $D_{0}(\alpha)$ at the end-points  $\alpha= \pm 1$
has the same nature as the vanishing 
of the DDs at the sides of the support rhombus:
both result from  a faster than perturbative decrease 
of the $k$-integrand at large quark virtualities.

\begin{figure}[h]
\begin{center} 
\includegraphics[scale=0.4]{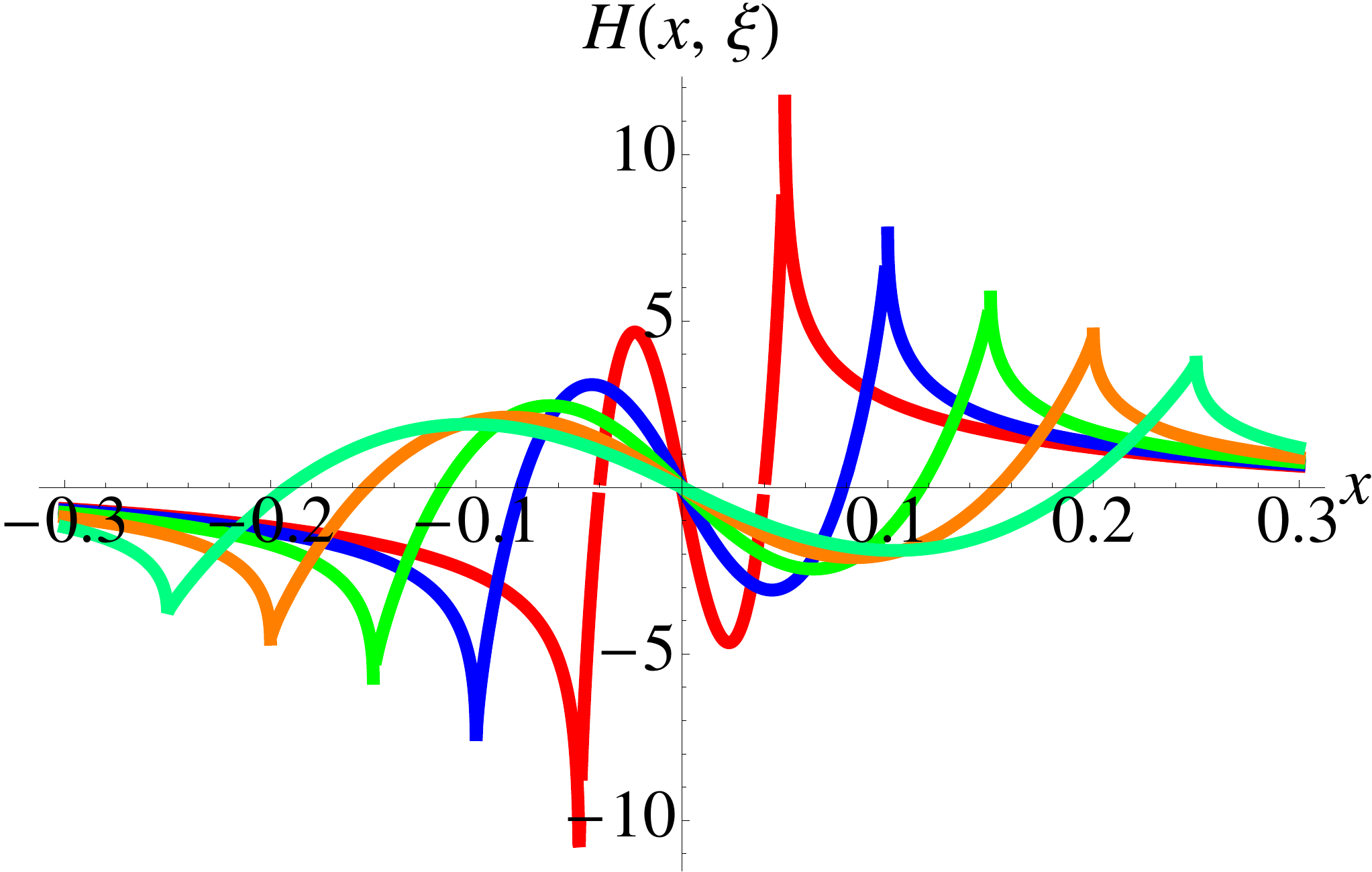} 
\end{center}
\caption{Model singlet GPD $H_S(x,\xi)$  with $N=1$ DD profile 
for $a=0.5$  and $\xi = 0.05, 0.1, 0.15,
0.2, 0.25$.}
\label{HN1}
\end{figure}

Turning   now to the  $|x|<\xi $  region, we use  Eq.~(\ref{eq_x<xi1})
to   represent  the relevant 
 term for the $N=1$  profile  as
\begin{align}
H  (x, \xi)|_{|x|<\xi}  = &\frac34\,  \frac{x}{\xi}  
\left [ \frac{1}{\xi^2}   \int_{0}^{\beta_2} 
  \frac{d\beta}{\beta^{a}}  (2x-\beta)
+
  \int_{0}^{\beta_2} 
  \frac{d\beta}{\beta^{a}} 
\left \{ 1- \frac{x^2}{\xi^2 (1-\beta)^2} \right \} 
(\beta-2)
 - 
    \int_{\beta_2}^{1-|x|/\xi} 
  \frac{d\beta}{\beta^{a+1}} 
\left \{ 1- \frac{x^2}{\xi^2 (1-\beta)^2}\right \} \right ]
\label{eq2_x<xi}  \  .
\end{align}
Note that as  far as $|x|$ is  strictly  less than $\xi$,  the profile function
does not   vanish at the  singularity  point  $\beta=0$.
The   mechanism of   softening singularity  to $1/\beta^a$  strength
is  now provided by  the $1/\sigma$ subtraction  term of the original dispersion relation.

To get a  model for singlet GPDs, one should 
take the  antisymmetric combination
\begin{align}
H_S(x,\xi) = H(x,\xi)   - H(-x,\xi)  \ .
\end{align}
The resulting GPDs are shown in Fig.~\ref{GPD1}.
For positive  $x$, they are  peaking  at $x=\xi$.
The functions $H_S (x, \xi)$ in this model  are continuous at
$x= \pm \xi$, but the   derivative $dH_S (x,\xi)/dx$ is  discontinuous
at these points.

\subsubsection{$N=2$ Profile}

Let us now take the $N=2$ profile function
\begin{align}
 h_2 (\beta, \alpha) = \frac{15}{16} \, \frac{[(1-\beta)^2 - \alpha^2]^2}{(1- \beta)^5}
\end{align}
and the same model forward distribution 
\begin{align}
 f_a (\beta)= \frac{(1-\beta)^3}{\beta^a}  \ .
\end{align}
For $x>\xi$ this gives
\begin{align}
H  (x, \xi)|_{x>\xi}  = &\frac{15}{16}  \,  \frac{x}{\xi}    \int_{x_1}^{x_2} 
  \frac{d\beta}{\beta^{a+1}(1-\beta)^2}
\left [ (1-\beta)^2 - \left (\frac{x-\beta}{\xi} \right )^2 \right ]^2  \  . 
\label{eq_x>xi2} 
\end{align}
Evidently,  the $N=2$  profile gives a  $\sim \beta^2$ suppression,
and $H(\xi,\xi)$  is  finite as  far as $a<2$.

Again, using Eq. (\ref{eq_x<xi1}), the $|x|<\xi $ term  can  be represented 
 in the form
\begin{align}
H  (x, \xi)|_{|x|<\xi}  = &\frac{15}{16}\,  \frac{x}{\xi^5}    \int_{0}^{x_2} 
  \frac{d\beta}{\beta^{a}}  (2x-\beta)\left [\frac{x^2+(x-\beta)^2}{(1-\beta)^2}-2\xi^2 \right ]
+\frac{15}{16}\,  \frac{x}{\xi}    \int_{0}^{x_2} 
  \frac{d\beta}{\beta^{a}} 
\left [ 1- \frac{x^2}{\xi^2 (1-\beta)^2} \right ]^2
(\beta-2) \nonumber \\
&- \frac{15}{16}\,  \frac{x}{\xi}    \int_{x_2}^{1-|x|/\xi} 
  \frac{d\beta}{\beta^{a+1}} 
\left [ 1- \frac{x^2}{\xi^2 (1-\beta)^2}\right ]^2 
\label{eq3_x<xi} 
\end{align}
explicitly showing the cancellation of the $1/\beta^{a+1}$ 
singularity.

The resulting GPDs are shown in Fig.~\ref{GPDN2}.
For positive  $x$, they are  peaking at points  close to  $x=\xi$.
 In the model  with $N=2$  profile,
both the functions $H_S (x, \xi)$  and their   derivatives $dH_S (x,\xi)/dx$ are continuous at
$x= \pm \xi$. 

\begin{figure}[h]
\begin{center} 
\includegraphics[scale=0.3]{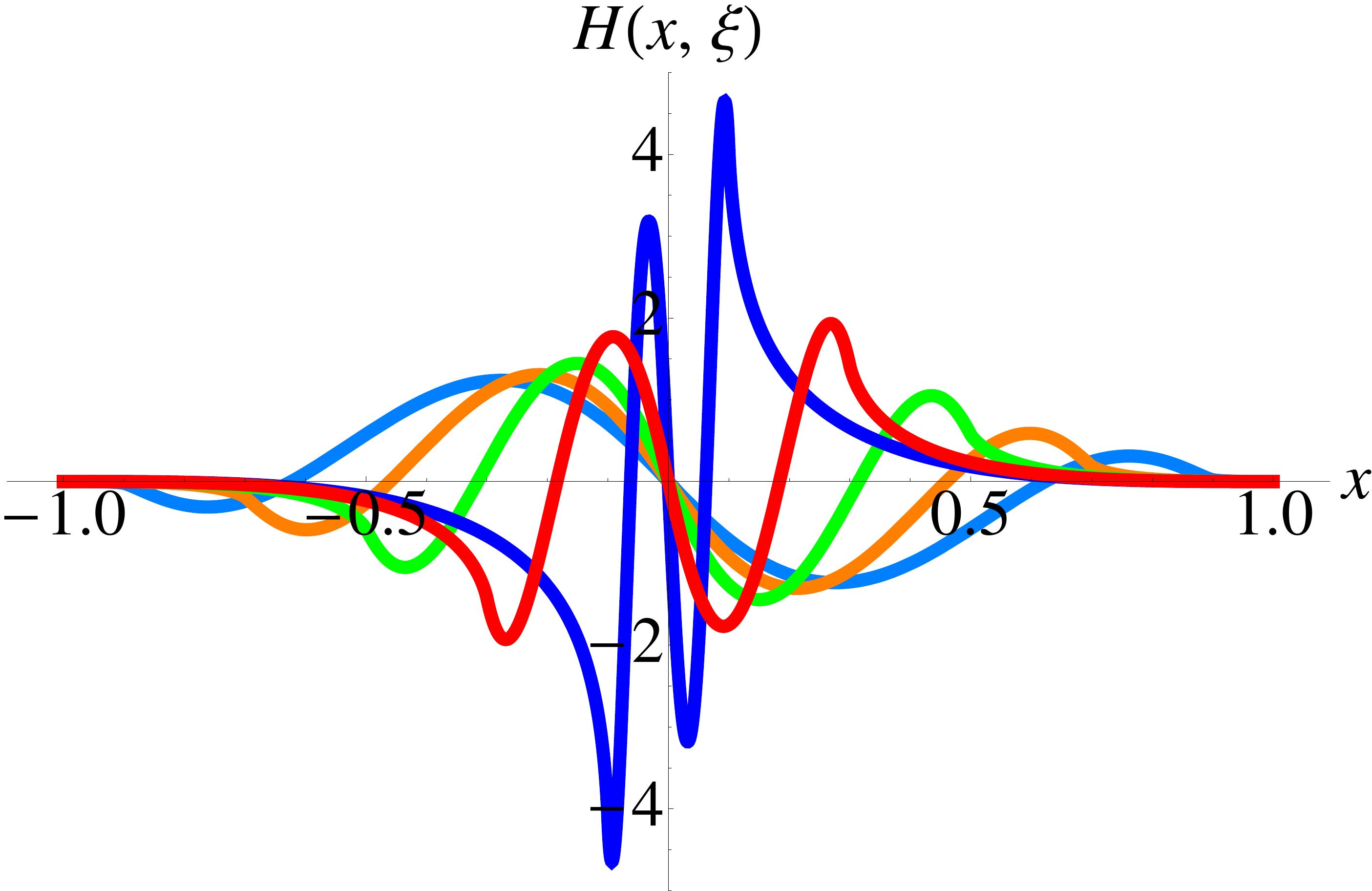} 
\end{center}
\caption{Model singlet GPD $H(x,\xi)$ with $N=2$ profile  for $a=0.5$  and $\xi = 0.1, 0.3, 0.5,
0.7, 0.9$.}
\label{GPDN2}
\end{figure}

\subsection{$D$-term} 

The $\delta (\beta)$ subtraction  term in the regularized DD  (extended now onto the whole
support rhombus)
\begin{align}
f^{\rm  reg} (\beta, \alpha)  =   f(\beta,\alpha) -  \delta (\beta)  
 \int_{-1+|\alpha|} ^{1  -|\alpha|} d \gamma \,  \frac{f (\gamma, \alpha)}
{(1-\gamma)^2} 
\label{eq_DDAreg2} 
\end{align}
softens the singularity of $f(\beta,\alpha)$  for $\beta=0$,
but it  does not convert $ f(\beta,\alpha)$ into a ``plus distribution'' 
$[f (\beta,\alpha)]_+$  whose integral over  $\beta$ vanishes.
Thus, $f^{\rm  reg} (\beta, \alpha) $  contains a  nonzero  $D$-term  contribution 
\begin{align}
D(\alpha) = &  \alpha \int_{-1+|\alpha|} ^{1  -|\alpha|} f^{\rm  reg} (\beta,\alpha) \, d\beta  
= \alpha  \int_{-1+|\alpha|}^{1-|\alpha|} d \beta \,    \frac{f(\beta)}{\beta}
 \, h(\beta,\alpha)
\, 
   \left \{  1- 
 \frac{1}{(1-|\beta|)^2}  
 \right \}
 =2 \alpha  \int_{0}^{1-|\alpha|} d \beta \,  
{f(\beta)} \, h(\beta,\alpha)
\, 
 \frac{\beta-2}{(1-\beta)^2} \  .
\label{eq_Dterm} 
\end{align}
Taking the same model forward distribution 
\mbox{$
 f (\beta)= {(1-\beta)^3}/{\beta^a}
$}
and $N=1$  profile function 
gives
\begin{align}
D^{\{N=1\}}(\alpha) =\frac32 \, \alpha  \int_{0}^{1-|\alpha|} \frac{d \beta}{\beta^a }  \,  
\left [1 - \frac{\alpha^2}{(1-\beta)^2} \right ]
\, 
(\beta-2) \  .
\label{eq_Dterm_1} 
\end{align}

A similar expression  for the $D$-term is  obtained in the $N=2$ 
profile model:
\begin{align}
D^{\{N=2\}}(\alpha) =\frac{15}{8} \, \alpha  \int_{0}^{1-|\alpha|} \frac{d \beta}{\beta^a }  \,  
\left [1 - \frac{\alpha^2}{(1-\beta)^2} \right ]^2
\, 
(\beta-2) \  .
\label{eq_Dterm_2} 
\end{align}
\end{widetext}
As one can  see in Fig. \ref{dt1and2}, the two curves are rather close to each other.

\begin{figure}[h]
\begin{center} 
\includegraphics[scale=0.16]{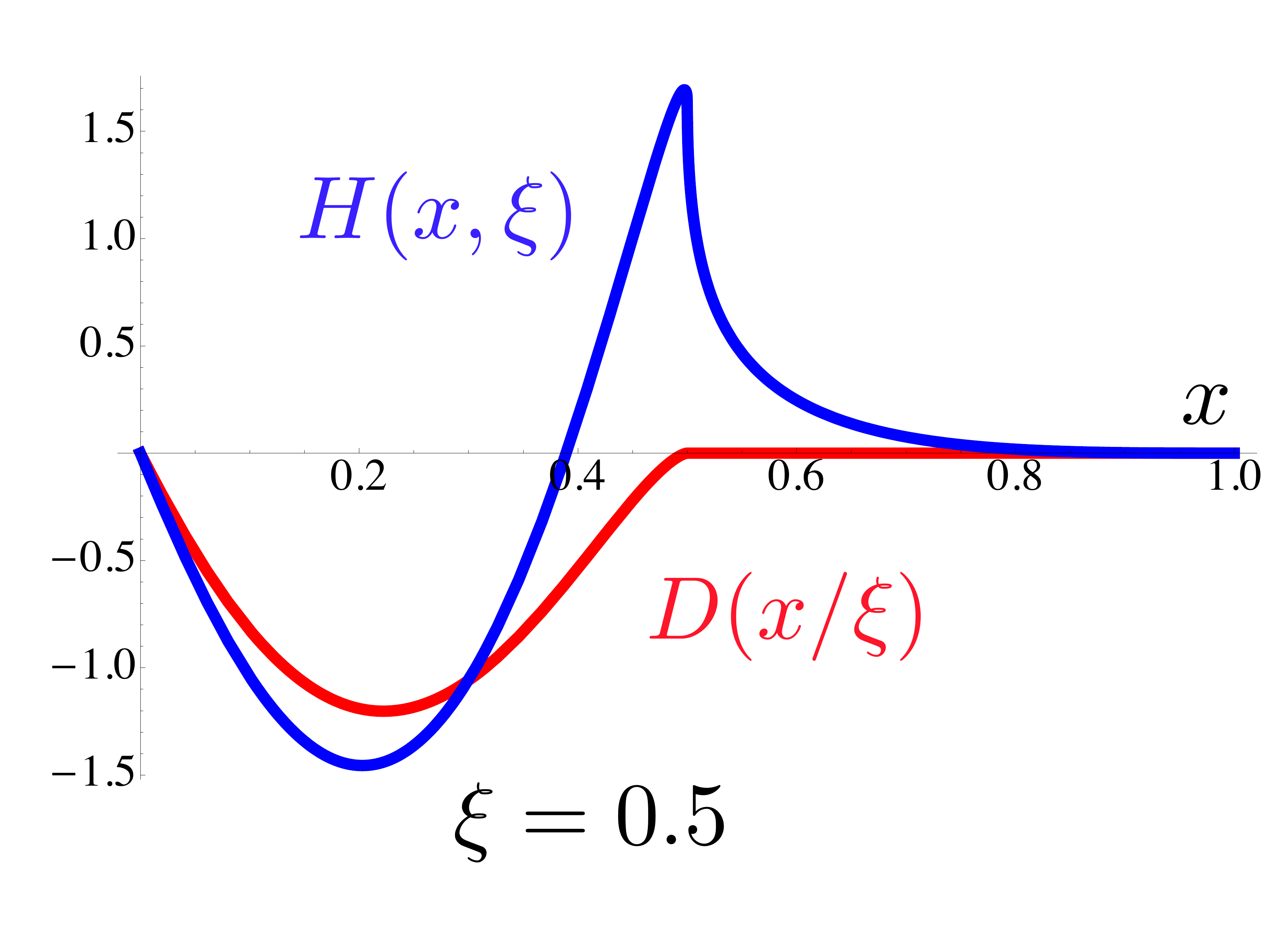} 
\end{center}
\vspace{-10mm}
\caption{GPD  $H(x,\xi)$  and $D$-term $D(x/\xi)$ for $\xi=0.5$ and positive $x$.}
\label{htvsdt}
\end{figure}

The comparison of the total GPD $H(x,\xi)$ and its  \mbox{$D$-term}  
part  is shown in Fig.~\ref{htvsdt}.  
\begin{figure}[t]
\begin{center} 
\includegraphics[scale=0.16]{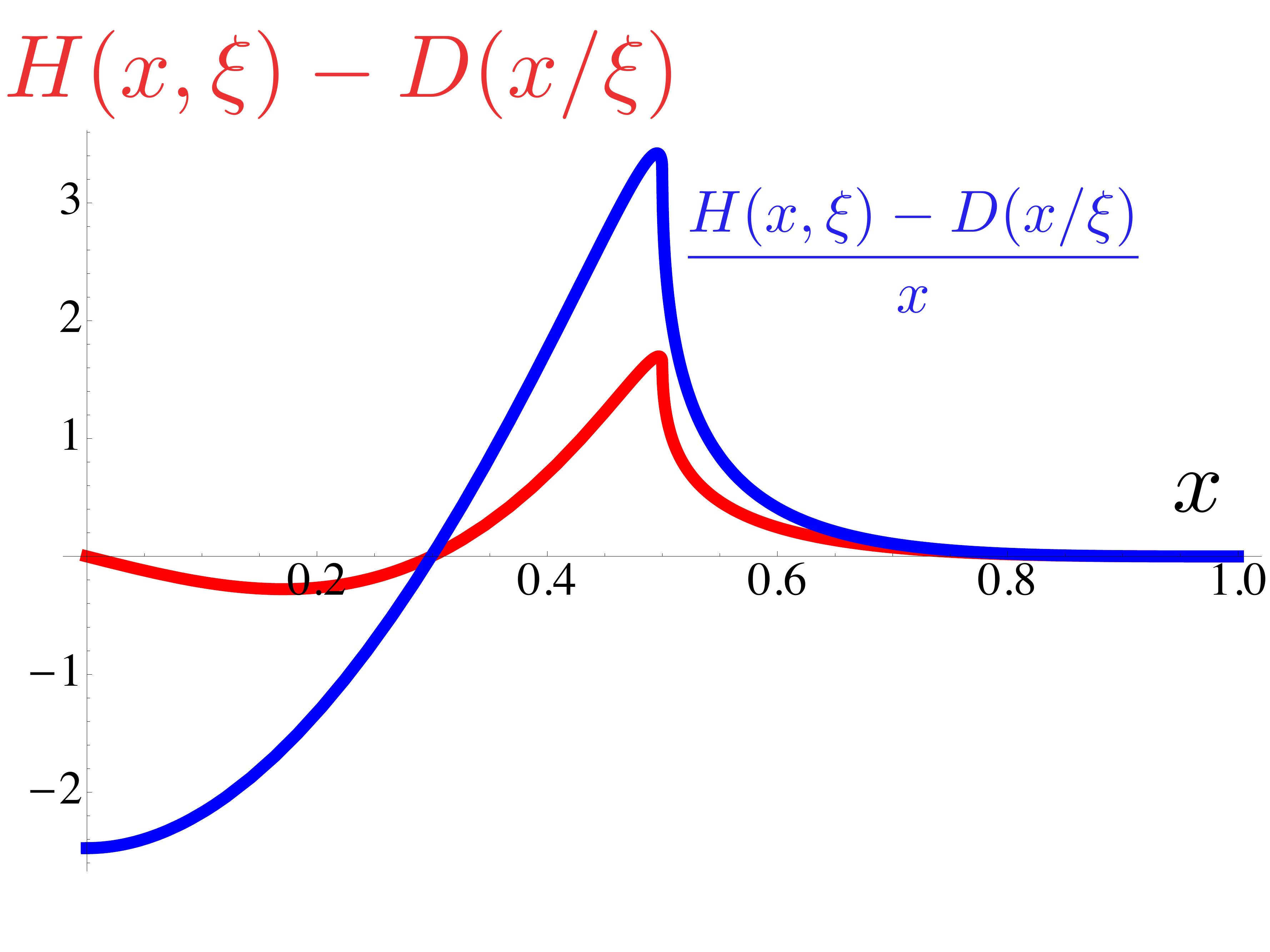} 
\end{center}
\vspace{-10mm}
\caption{Difference between GPD  $H(x,\xi)$  and $D$-term 
$D(x/\xi)$ in the case of the $N=1$ profile
for $\xi=0.5$ and positive $x$. The same function divided by $x$ is
also shown.}
\label{htminusdt}
\end{figure}
The difference between GPD  $H(x,\xi)$  and $D$-term $D(x/\xi)$ 
corresponds to the term $H_+(x,\xi)$  obtained from  the ``plus''  part 
$[ f(\beta,\alpha)]_+$  of DD. The  shape  of the difference  for $\xi=0.5$ is  shown in Fig.~\ref{htminusdt}.
Note that, despite the fact that the forward distribution
in this  model is positive,  
there is a region, where 
the contribution to $H(x,\xi)$  coming from $[ f(\beta,\alpha)]_+$
is  negative.
This is due to the  $\delta (\beta)$  subtraction term contained in $[ f(\beta,\alpha)]_+$.
Also   shown is the ratio $H_+(x,\xi)/x$. 
Looking at the figure, one may  suspect that the $x$-integral of
$H_+(x,\xi)/x$   vanishes.
In the next section, we show that this, indeed,  is the case.

\section{GPD sum rules}

\subsection{Formulation} 

The $D$-term  determines the subtraction
constant in the dispersion relation for the DVCS amplitude 
\cite{Teryaev:2005uj,Anikin:2007yh,Kumericki:2007sa,Diehl:2007jb,Teryaev:2010zz}.
In particular,   it was shown \cite{Anikin:2007yh}
that the  original expression  for the real part of the DVCS amplitude
involving  $H(x,\xi)$,   and  the dispersion integral involving 
$H(x,x)$ differ  by a  constant $\Delta$ given  by the integral of the $D$-term function
$D(\alpha)$:
\begin{align}
 P\int_{-1}^1  \frac{H (x, x) - H(x,\xi)}{x-\xi}\,  dx  =  \Delta  \equiv 
  \int_{-1}^1  \frac{D(\alpha)}{1-\alpha} \, d \alpha \ .
 \label{eq_SRG} 
\end{align}
 Here, $P$  denotes the principal value prescription. 
In Ref.~\cite{Anikin:2007yh}, this relation 
was  derived using polynomiality properties
of GPDs.  It was also pointed out there that it can  be obtained
by incorporating representation of GPDs  in the two-DD  formalism
(which is basically again the use of the polynomiality).

Taking $\xi=0$,  one formally arrives at the 
sum  rule 
\begin{align}
 \int_{-1}^1  \frac{H (x, x) - H(x,0)}{x}\,  dx  = 
  \int_{-1}^1  \frac{D(\alpha)}{1-\alpha} \, d \alpha \ .
 \label{eq_SR} 
\end{align}
Since    both
$H(x,0)/x$  and $H(x,x)/x$ are even functions
of $x$,    their singularities for $x=0$
cannot  be regularized by the principle value prescription.
Moreover, there are no indications
that singularities of these two functions may cancel
each other. On the contrary, as  emphasized in Ref.~\cite{Polyakov:2007rv},  there
are arguments that the ratio $H(x,x)/H(x,0)$  does not tend to 1
for small $x$.

The  solution given in Refs.~\cite{Kumericki:2007sa,Kumericki:2008di,Polyakov:2008aa}
is   based  on the analytic  regularization of the $x$-integral.
Namely, it  is assumed  that 
the positive  Mellin moments   (or   conformal moments,  see, e.g., \cite{Kumericki:2009uq}) 
\begin{align}
 \Phi (j) \equiv \int_{-1}^1 x^j[H (x, x) - H(x,0)] \,  dx 
 \label{eq_Phi} 
\end{align}
can  be analytically   continued to 
the point $j=-1$. The result of such 
a procedure is  equivalent to
analytic regularization of the $x$-integral. 
However, the  required  analyticity 
properties of $\Phi (j)$ may   be  violated
by   singular or    ``invisible'' 
terms (cf. \cite{Kumericki:2007sa})
 in the integrand of Eq.(\ref{eq_Phi})  (e.g., $x \delta (x)$  gives a  non-analytic $\delta_{j,-1}$ 
 contribution into $\Phi (j)$).
In  the   model construction described 
above,  singular  terms explicitly  appear as  a result 
of  subtractions  in  the dispersion  relation, 
so  our  intention is  to develop  a  less restrictive  approach 
to this problem.

Below,  we give a  derivation of the  sum rule  (\ref{eq_SR} )
based on    separation (\ref {Plussingle})   of the DDs 
into the ``plus'' part and the $D$-term.
No  assumptions about smoothness will be made.
In  fact,  the key  element  of the derivation is  that 
$H(x,x)/x$ should be treated as a 
 (mathematical) distribution at the point  $x=0$
rather than  a function. The same applies to $H(x,0)/x$.

\subsection{Ingredients} 

To begin with, we remind    the basic formulas: the expression 
\begin{align}
 \frac{H(x,\xi) }{x}  =
 \int_{\Omega}    f(\beta,\alpha) \,  \delta (x - \beta -\xi \alpha)  \, d\beta \, d\alpha  
\label{GPDf2}
\end{align}
producing GPDs from DDs  and the decomposition of DD
\begin{align}
  f(\beta,\alpha)= [f(\beta,\alpha)]_+  +\delta (\beta) \frac{D(\alpha)}{\alpha}  
\end{align}
into  the  ``plus'' part   given by
\begin{align}
 [f(\beta,\alpha)]_+=   f(\beta,\alpha) - \delta (\beta)  
  \int_{-1+|\alpha|} ^{1  -|\alpha|} f(\gamma,\alpha) \, d\gamma
\end{align}
and the $D$-term part   $ \delta (\beta) {D(\alpha)}/{\alpha}$.

Correspondingly, we  split GPD into the part coming from 
the ``plus'' part  of DD 
\begin{align}
 \frac{H_+(x,\xi) }{x}  \equiv & 
 \int_{\Omega} f(\beta,\alpha)   \,\biggl [   \delta (x - \beta -\xi \alpha) 
 - \delta (x - \xi \alpha )  \biggr ]
  \, d\beta \, d\alpha  
\label{GPDfplus}
\end{align}
and  that  generated by  the $D$-term
\begin{align}
 \frac{H_D(x,\xi) }{x}  \equiv & 
 \int_{-1} ^{1 }  \frac{D(\alpha)}{\alpha}   \delta (x - \xi \alpha) 
\, d\alpha  \  .
\label{GPDD}
\end{align}
The latter integral gives an  explicit expression 
\begin{align}
H_D(x,\xi) =  {\rm sign}(\xi)\,   \theta (|x|< |\xi|)\, D(x/\xi)\ ,
\label{Dexpl}
\end{align}
but, as we will see,  it is instructive to  use the integral representation  as well.
Another important relation
\begin{align}
 \frac{H_D(x,0) }{x}  = &  \delta (x) 
 \int_{-1} ^{1 }  \frac{D(\alpha)}{\alpha}  
\, d\alpha 
\label{GPDDxi0}
\end{align}
 is  obtained by taking $\xi=0$.

Note,  that  if we take $x=\xi$, Eq.~(\ref{Dexpl})  gives
\begin{align}
H_D(x,x) ={\rm sign}(x) \,  D(1) \ .
\label{D1}
\end{align}
If $D(1) \neq 0$, then the integral for $\Delta$  in (\ref{eq_SRG})
diverges. 
However, as argued in the previous section,  
 \mbox{$D(1) = 0$}
for  models with faster-than-perturbative 
decrease of the hadron-parton amplitude at large quark virtualities.
Thus, we assume that $D(1)=0$,  and furthermore that 
the integral of $D(\alpha)/(1-\alpha)$ converges.  Then Eq.~(\ref{GPDD})  gives
\begin{align}
 \frac{H_D(x,x) }{x}  =&  \delta (x ) 
 \int_{-1} ^{1 }  \frac{D(\alpha)}{\alpha(1-\alpha)}  
\, d\alpha  \  .
\label{GPDDxx}
\end{align}
The important point is that if we would use this  formula  to write  an expression 
for $H_D(x,x)$  itself, we would get
$x \delta (x)$ on the r.h.s., which should  be treated as zero
for   integration with   functions finite   for $x=0$,
since the coefficient  given by the $\alpha$-integral 
is  also finite. Thus,  the  scenario with $D(1)=0$ is self-consistent.

Note that
both $H_{D}(x,0)/x$  and $H_{D}(x,x)/x$  are 
proportional to  $\delta (x)$,
with the coefficients 
given  by integrals of $D(\alpha)$. 
This means that, unlike  the functions $H(x,0)$  and $H(x,x)$,
which,   for $x \neq 0$,   are insensitive  to  changes of
$D(\alpha)$  in the $\delta (\beta)  D(\alpha)/\alpha $  term, 
 the (mathematical)  distributions $H(x,0)/x$  and $H(x,x)/x$
contain information    about such a  $D$-term.

Our   next step is to study contributions
from different parts of the GPDs  involved in the  sum rule
(\ref{eq_SR}).

\subsection{``Secondary'' sum rule}

\subsubsection{``Plus'' part} 

\paragraph{Forward function}

\begin{figure}[h]
\begin{center} 
\includegraphics[scale=0.2]{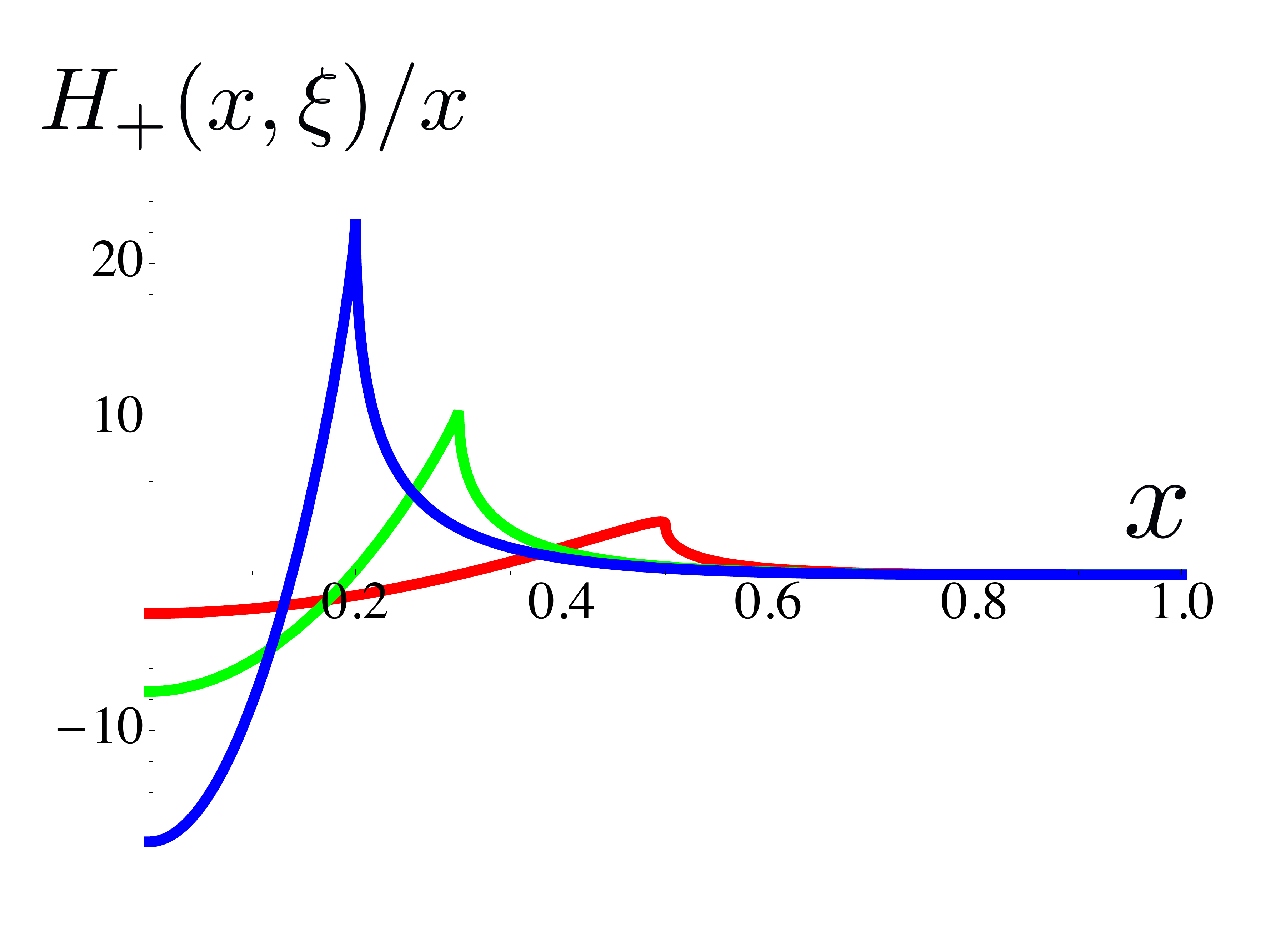} 
\end{center}
\vspace{-10mm}
\caption{Function  $H_+(x,\xi)/x$   in the    model with  the $N=1$ profile
for $\xi=0.2,0.3,0.5$ and positive $x$.}
\label{Hplus_x}
\end{figure}

One  can easily  see from Eq.~(\ref{GPDfplus})  that
\begin{align}
 \int_{-1}^1  \frac{H_+(x,\xi) }{x}  \,  dx  =0 \label{eq_SR+} 
\end{align}
for any $\xi$, including $\xi=0$.
Since the integrand is an even function of $x$,  the vanishing of this integral 
means that 
 we also have
\begin{align}
 \int_{0}^1  \frac{H_+(x,\xi) }{x}  \,  dx  =0 \label{eq_SR++}   \  .
\end{align}
Thus, 
 $H_+(x,\xi) $ should be negative in some part 
of the central region, and this negative  contribution
should exactly compensate the contribution from the regions, 
where $H_+(x,\xi) $ is positive.
 In other words,  on the $(0,1)$  interval,  $H_+(x,\xi) /x$
 has the same property as a  ``plus distribution'' with  respect to $x$.
 Note, that this  does not mean that  $H_+(x,\xi) /x$ necessarily contains 
singular functions like $\delta (x)$.
 For  finite $\xi$, the function $H_+(x,\xi)/x $
 is pretty   regular for all $x$ values  (see Fig.\ref{Hplus_x}).
 The negative $\delta (x)$   function appears only in the $\xi$=0 limit, i.e.
 \begin{align}
   \frac{H_+(x,0) }{x}   =  \frac{f(x) }{x}  -\delta (x)  \int_{-1}^1    \frac{f(y) }{y}\,  dy 
   \equiv \left [  \frac{f(x) }{x} \right ]_+
 \  . 
 \label{eq_SRB+xx} 
\end{align}
 (Here, it was  taken into account that $H_+(x,0)$  coincides  with the forward
 distribution $f(x)$ for $x\neq 0$).

\paragraph{Border function}

For the integral involving the border function, we get
\begin{align}
 \int_{-1}^1  \frac{H_+(x,x) }{x}  \, & dx  =
 \int_{-1}^1  dx  \, 
 \int_{\Omega}  
  \, d\beta \, d\alpha \ 
{f(\beta,\alpha)}  \nonumber \\ 
 \times &
  \, \left  [   \delta \left (x (1-\alpha )- \beta \right  ) 
 - \delta (x (1-\alpha))  \right ]  \  .
 \label{eq_SRB} 
\end{align}
Noting that  equation $\alpha=1$ is satisfied  in one point on the 
support region $\Omega$ only, namely, in the upper corner of the rhombus,
we may treat $ \delta (x (1-\alpha))$  as $\delta (x)/ (1-\alpha)$
to get 
\begin{align}
 \int_{-1}^1  \frac{H_+(x,x) }{x}  \,  dx  =
 & \int_{-1}^1  dx  \, 
 \int_{\Omega}  
  \, d\beta \, d\alpha \ 
 \frac{f(\beta,\alpha)}{1-\alpha}   \nonumber \\ & 
 \times 
  \, \left  [   \delta \left (x - \frac{\beta}{1-\alpha} \right  ) 
 - \delta (x )  \right ]  \  .
 \label{eq_SRB2} 
\end{align}
Since $|\beta/(1-\alpha)| \leq 1$,   the first $\delta$-function
always works. As a result,  
the integrals  coming from  the two delta-functions cancel each other,
and we have
\begin{align}
 \int_{-1}^1  \frac{H_+(x,x) }{x}  \,  dx  =0 
 \  ,
 \label{eq_SRB+} 
\end{align}
just like for ${H_+(x,\xi) }/{x} $.     Unlike $H_+(x, \xi)$,  however,
the combination $H_{+}(x,x)/x$ 
 explicitly  contains   the $\delta (x)$  subtraction term,  i.e. it 
is   a genuine ``plus distribution''  with  
respect to $x$: 
\begin{align}
   \frac{H_+(x,x) }{x}   = &  \frac{H(x,x) }{x}  -\delta (x)  \int_{-1}^1    \frac{H(y,y) }{y}\,  dy 
 \nonumber \\ &   \equiv \left ( \frac{H(x,x) }{x} \right )_{+}  \ .
 \label{eq_SRB+xx2} 
\end{align}

Summarizing,  the ``plus''  parts  of both  functions
entering into the sum rule (\ref{eq_SR}) separately
produce vanishing   contributions into the $x$-integral.
Furthermore,  these   zero contributions 
are due to the   fact that $H_+(x,0)/x$  and $H_+(x,x)/x$
are ``plus  distributions'',
which results in  zero  integrals irrespectively 
of the form of the forward distribution $f(x)$ and the 
border function $H(x,x)$.   

\subsubsection{$D$-part}

Let us now turn to the $D$-parts. 
First, we have 
\begin{align}
 \int_{-1}^1  \frac{H_D(x,\xi) }{x}  \,  dx 
 =& 
 \int_{-1} ^{1 }  \frac{D(\alpha)}{\alpha}   
\, d\alpha  
 \label{Hdxi1} 
\end{align}
for any fixed $\xi$, including $\xi=0$.
 This  result  may be  obtained 
by  integrating  over $x$ the $\delta (x -\xi \alpha)$ factor 
in the integral representation (\ref{GPDD}).
For non-vanising $\xi$, one can also use  
  Eq. (\ref{Dexpl})  in the $x$-integral and then  change
the integration variable through $x=\alpha \xi$.

For the  integral involving the border function,  we use 
Eq.~(\ref{GPDDxx}), which gives
\begin{align}
 \int_{-1}^1  \frac{H_D(x,x) }{x}  \,  dx  {=}
 & 
 \int_{-1} ^{1 }  \frac{D(\alpha)}{\alpha(1-\alpha)}   
\, d\alpha  \    .
 \label{eq_SRDDxx} 
\end{align}
As a result,
\begin{align}
 \int_{-1}^1  \frac{H_D (x, x) }{x}\,  dx
 - \int_{-1}^1  \frac{ H_D(x,0)}{x}\,  dx  {=}  
  \int_{-1}^1  \frac{D(\alpha)}{1-\alpha} \, d \alpha \ .
 \label{eq_SRDD} 
\end{align}
Combining this  outcome with  zero contributions
from the ``plus'' parts,  one 
obtains  the sum rule (\ref{eq_SR}). 

Thus, our  construction   confirms the sum rule. 
But  our  derivation   shows that 
the ``plus''  parts  of both terms 
simply do not contribute
to the sum rule   whatever 
 the shapes of $f(x)$  and $H(x,x)$  are.  
Only  the $D$-parts  contribute,
so  there is no surprise that
the  net result  can be expressed 
in terms of $D(\alpha)$.

An  essential point is that
both $H_{D}(x,0)/x$  and $H_{D}(x,x)/x$  are 
proportional to the $\delta (x)$-function,
with the coefficients 
given  by integrals of the $D$-term function
$D(\alpha)$.
In this sense,  $H(x,0)/x$  and $H(x,x)/x$
``know''  about the $D$-term.

A simple consequence is that all $x^{j}$  moments
of $H_{D}(x,0)$  and $H_{D}(x,x)$
with $j\geq 0$ vanish and  one  cannot get the
$D$-part of the sum rule  (\ref{eq_SR}) by 
an  analytic  continuation 
of the $x^{j}$  moments 
of  $H_D(x,0)$  and $H_D(x,x)$ to $j=-1$,
i.e.,  using the procedure 
of 
Refs.\cite{Kumericki:2007sa,Kumericki:2008di,Polyakov:2008aa}. 
In fact, $x^{j}$  moments of $H_{D}(x,0)$  and $H_{D}(x,x)$
are proportional to the Kronecker delta function 
$\delta_{j,-1}$.

\subsubsection{Formal   derivation  and  need for renormalization}

Since $H(x,0)/x$ is  given  by integrating the DD $f(\beta,\alpha)$ over $\alpha$ 
along vertical lines $\beta=x$, a subsequent integration over all $x$ 
gives  DD $f(\beta,\alpha)$  integrated over the whole rhombus:
\begin{align}
 \int_{-1}^1  \frac{H (x,0) }{x}  \,  dx   & {=}  
 \int_{-1} ^{1 }  dx   \int_{\Omega}
  \, d\beta \, d\alpha    f(\beta,\alpha)  \delta (x- \beta)  \nonumber \\ &
 =    \int_{\Omega} f(\beta,\alpha)   \,
  \, d\beta \, d\alpha      =    \int_{-1}^1  \frac{D(\alpha)}{\alpha} \, d \alpha 
 \    .
 \label{eq_SRDDxx2} 
\end{align}
On   the last  step, we used that the $\beta$-integral of  $f(\beta,\alpha)$   formally 
gives $D(\alpha)/\alpha$.  
However, if $f(\beta,\alpha) \sim 1/\beta^{{1+a}}$, being even in $\beta$, one  needs a regularization
for the \mbox{$\beta$-integral.} The ``DD+D'' separation (\ref{GPDf2}), as we have seen,
provides such a regularization. It works like a renormalization:
the divergent  integral formally giving the $D$-term 
is subtracted from the
``bare'' DD, and substituted by a  finite ``observable'' function $D(\alpha)/\alpha$.

In a similar way, we can treat the second integral:
\begin{align}
 \int_{-1}^1  \frac{H (x,x) }{x}  \,  dx    &{=}  
 \int_{-1} ^{1 }  dx   \int_{\Omega} 
  \, d\beta \, d\alpha    f(\beta,\alpha)  \delta (x- \beta -x \alpha)  \nonumber \\ 
 =  &  \int_{\Omega}  \frac{f(\beta,\alpha)}{1-\alpha}   \,
  \, d\beta \, d\alpha      =    \int_{-1}^1  \frac{D(\alpha)}{\alpha (1-\alpha)} \, d \alpha 
 \    .
 \label{eq_SRDDxx3} 
\end{align}
Again, the last step 
requires a subtraction of the infinite part of the $\beta$-integral. 

The advantage of using the  ``DD+D'' separation  as a  renormalization
prescription  is that it is  applied directly to the DD.   Hence,
it is universal, and  may  be used for other  integrals involving
$f(\beta, \alpha)$.

\subsubsection{Comparison of the ``plus'' prescription and analytic 
regularization}

Another  possibility to renormalize the $\beta$-integral  for a singular DD
is to use the analytic regularization 
employed in Refs.\cite{Kumericki:2007sa,Kumericki:2008di,Polyakov:2008aa}.
It works as follows. If we need to integrate a function like 
$\lambda (x)/x^{a+1}$  with $\lambda (x)$  being finite and nonzero
for \mbox{$x=0$,} we subtract from $\lambda (x)$ as many terms 
of its Taylor expansion as needed to  remove the  divergence
\begin{align}
 &\int_{(0)}^{y}  \frac{ \lambda (x)}{x^{a+1}} \, dx 
 =   \int_0^y  \, dx  \frac{\lambda (x) -\lambda (0) -x \lambda^{\prime} (0) -\ldots}{x^{a+1}}
 \nonumber   \\ & +   \lambda (0)  \int_{(0)}^{y}  \frac{dx }{x^{a+1}} +
    \lambda^{\prime} (0)   \int_{(0)}^{y}  \frac{dx}{x^{a}} +\ldots \
    ,  
 \label{analyt0} 
 \end{align}
and then treat the compensating integrals of $x^n/x^{a+1}$ as convergent,
substituting them by $y^{n-a}/(n-a)$.

So, let us  consider again 
 a DD which is  nonzero  for positive $\beta$ only and  has the form  
$$f(\beta, \alpha) = \frac{\lambda (\beta,\alpha)}{\beta^{a+1}} \, \theta (\beta+|\alpha| \leq  1) \, \theta (\beta \geq 0)$$
with $a<1$. Then the analytic regularization of its integral  
with some reference  function $\Phi (\beta)$ is defined by
\begin{align}
 &\int_{(0)}^{1-|\alpha|} \frac{  \Phi (\beta) \lambda (\beta,\alpha)}{\beta^{a+1}}
 \, d\beta \\ 
 =  &   \int_0^{1-|\alpha|}  \frac{ \Phi (\beta) \lambda (\beta,\alpha)-  \Phi (0) \lambda (0, \alpha)}{\beta^{a+1}}
 \, d\beta  - \frac{  \Phi (0) \lambda (0, \alpha)}{a (1-|\alpha|)^a}
 \    , \, \nonumber  
 \label{analyt} 
 \end{align}
 which may be rewritten as
 \begin{align}
 &\int_{(0)}^{1-|\alpha|} \Phi (\beta) \frac{  \lambda (\beta,\alpha)}{\beta^{a+1}}
 \, d\beta  \nonumber   \\ 
  & =
  \int_0^{1-|\alpha|}  [\Phi (\beta) -  \Phi (0)] \,  \frac{  \lambda (\beta,\alpha)}{\beta^{a+1}} 
 \, d\beta  
 \\ & + \Phi (0)  \left [ \int_0^{1-|\alpha|}  \frac{ \lambda (\beta,\alpha)- \lambda (0,\alpha)}{\beta^{a+1}}
 \, d\beta  
 - \frac{  \lambda (0, \alpha)}{a (1-|\alpha|)^a} \right  ]  \ . \nonumber  
   \label{analyt2} 
\end{align}
Now, the first contribution on the r.h.s. 
is generated by  the ``plus'' part of the DD,  while  the second one comes from   
a $D$-term. After adding the $\beta<0$ part of  the  DD, the $D$-term
${\cal D}(\alpha)/{\alpha}$  corresponding to the analytic regularization 
 is  given by 
 \begin{align}
 %  \label{anaD} 
 \frac{{\cal D}(\alpha)}{\alpha} = 2 \left [
 \int_0^{1-|\alpha|}  \frac{ \lambda (\beta,\alpha)- \lambda (0,\alpha)}{\beta^{a+1}}
 \, d\beta  
 - \frac{  \lambda (0, \alpha)}{a (1-|\alpha|)^a} \right  ] \ .
\end{align}
Thus, the analytic regularization  prescription unambiguously 
fixes the $D$ term, and in this sense it may be called the 
``analytic renormalization''.

In the model considered in the previous section,
we also obtained a  concrete result for the $D$-term.  
But the specific $D$-term contribution we obtained there 
came only  from the $\sigma$-integral part of the 
dispersion relation  for the hadron-parton scattering amplutude
subtracted at \mbox{$(P-k)^2=0$.}
As we pointed out, one should be  always  ready 
to add to it the $D_0$ term coming from the $T_0$ constant
 in the dispersion relation (\ref{DRel}).
In principle,  we had   no reasons  to require that $T_0=0$.
In  this sense, the $D$-term in that  model is  not  fixed.

On  the other   hand, the statement, 
that  $x^j$ moments of  $H(x,\xi)$
are analytic  functions of $j$,  does  not  explicitly  mention fixing any
subtraction constants:    it 
sounds   like a  general principle,
and  may  create an impression that there are
no ambiguities in the subtraction of the $\beta=0$
singularity.   However,  the analyticity assumption was not 
shown so far to be a consequence of general principles
of quantum field theory. 
Moreover, as mentioned in Ref.~\cite{SemenovTianShansky:2008mp},
 it 
is  not satisfied in the  nonlocal  chiral  soliton  model.  
Still, one  may  hope that  it is  valid in QCD. 

\begin{figure}[t]
\begin{center} 
\includegraphics[scale=0.2]{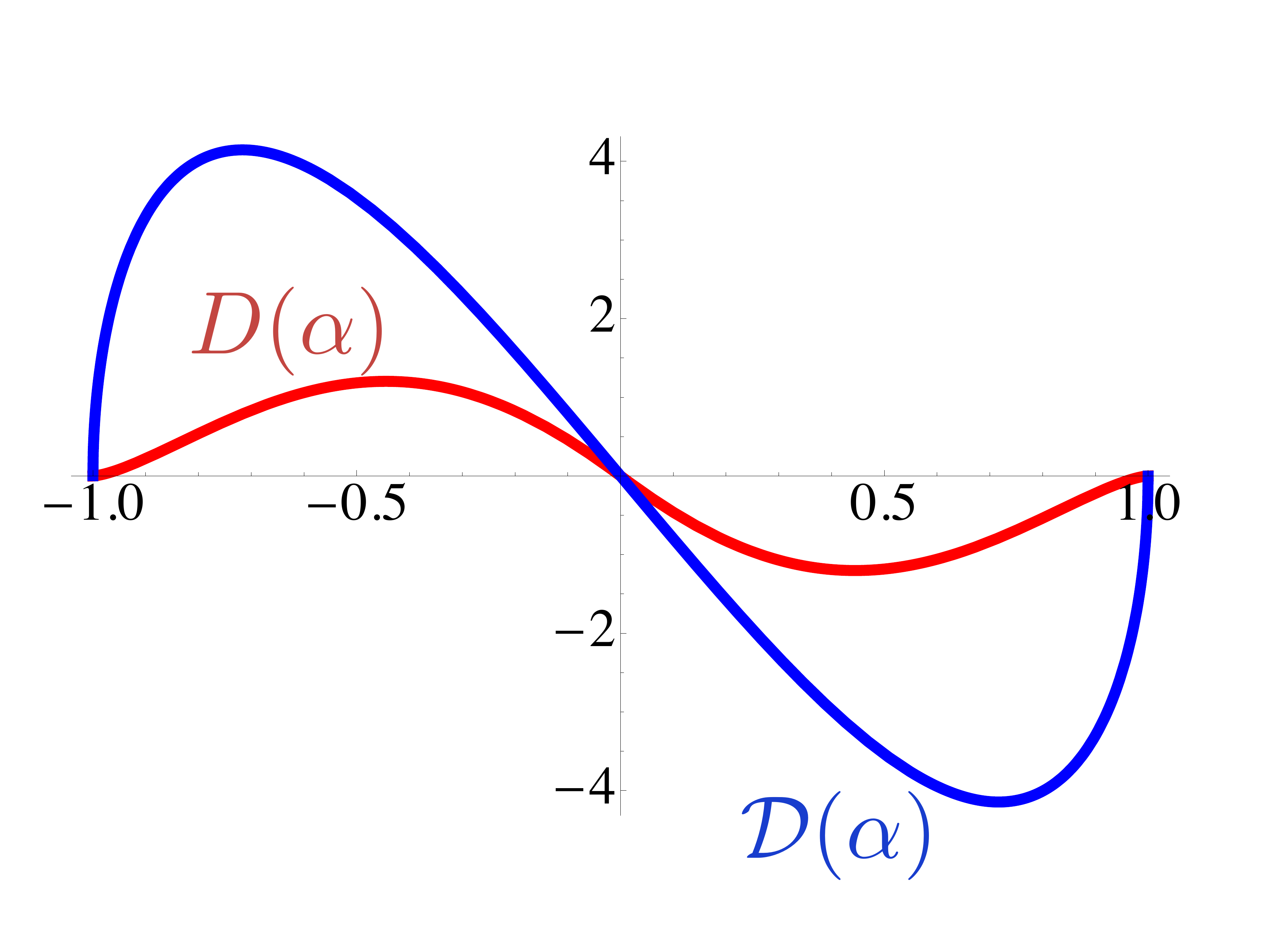} 
%\vspace{-1cm}
\caption{The  $D$-terms    in  the model with $N=1$  profile  and  $a=0.5$: 
 ${\cal D} (\alpha)$  was  obtained using analytic  regularization,  and 
 $D(\alpha)$ was obtained for $T_0=0$ in the model of the previous section.}
\label{danal}
\end{center}
\end{figure}

To see if  the  $T_0=0$  model 
of the previous section agrees with  the analyticity assumption,
we should  just check whether its 
$D$-term  is different  from 
that  obtained {\it via} analytic  renormalization.
In  particular, for the $N=1$ model,
we have
 \begin{align}
 \lambda (\beta,\alpha)= \frac34 \left [ (1-\beta)^2 -\alpha^2  \right ]    \ , 
%  \label{analytD} 
\end{align}
and, hence,
 \begin{align}
 \frac{{\cal D}(\alpha)}{\alpha} =  \frac32 \left [ 
 \frac{ (1- |\alpha|)^{2-a}}{2-a} \,  -2 \frac{(1- |\alpha|)^{1-a}}{1-a} \, - \frac{  1- \alpha^2}{a (1-|\alpha|)^a} \right ]  \,  .
%  \label{analytD2} 
\end{align}
In  Fig.\ref{danal},  we compare this result  (for $a=0.5$)  with the result
obtained by single subtraction in the dispersion relation (\ref{DRel}) with $T_0=0$.

Our  main point is that representing $H(x,\xi)$  as the sum 
$H_+(x,\xi)+ H_D(x,\xi)$ one   can  derive the GPD  sum rule (\ref{eq_SR})
without using the analyticity assumption.  But since 
our  derivation,  so to say, works   for any $D$-term, it also works  for the $D$-term   following  from the analyticity assumption.

\subsection{Generic sum rule}

 Finally, let us    apply the ``DD+D'' separation   to   the generic relation (\ref{eq_SRG}).

\subsubsection{``Plus''  part}

Representing 
\begin{align}
\frac1{x-\xi} = \frac1{x} + \frac{\xi}{(x-\xi) x}
\end{align}
and using Eqs.~(\ref{eq_SR++}), (\ref{eq_SRB+}), we have
\begin{align}  &
P \int_{-1}^1  \frac{H_+ (x, x) }{x-\xi}\,  dx  
 = 
 P \int_{-1}^1   \xi \, \frac{dx}{x-\xi}\,    
 \int_{\Omega} f(\beta,\alpha)   \,
  \, d\beta \, d\alpha   
  \nonumber \\ & \times
  \biggl [ \delta (x(1-\alpha) - \beta ) -   \delta (x(1-\alpha) )  
     \biggr ]
     \\ & 
     =   
P  \int_{\Omega} f(\beta,\alpha)   \,
  \, d\beta \, d\alpha   
  \biggl [ \frac{\xi}{\beta- \xi (1-\alpha) }+ \frac1{  (1-\alpha) } 
     \biggr ]  \nonumber  \  , 
 \label{eq_SRG2xx} 
\end{align}
and
\begin{align}  &
P \int_{-1}^1  \frac{ H_+(x,\xi)}{x-\xi}\,  dx   = 
  P \int_{-1}^1   \xi  \,  \frac{dx}{x-\xi}\,    
  \int_{\Omega} f(\beta,\alpha)   \,
  \, d\beta \, d\alpha   
  \nonumber \\ & \times
  \biggl [ \delta (x - \beta -\xi \alpha)  
 - \delta (x - \xi \alpha )  \biggr ]
 \\ &
  =    
P \int_{\Omega} f(\beta,\alpha)   \,
  \, d\beta \, d\alpha   
  \biggl [ \frac {\xi  }{\beta- \xi (1-\alpha) }+ \frac1{  (1-\alpha) } 
     \biggr ]   \ . \nonumber
 \label{eqxxi} 
\end{align}
Thus, seemingly different delta-functions have converted
$1/(x-\xi)$  into identical  expressions  
(cf. Ref.~\cite{Diehl:2007jb}, where a similar result was obtained 
for the $F_D$ part of the two-DD representation).
As a result,  
\begin{align}
 P\int_{-1}^1  \frac{H_+ (x, x) }{x-\xi}\,  dx - P\int_{-1}^1  \frac{ H_+(x,\xi)}{x-\xi}\,  dx  =  0  \  .
\end{align}
In this case,  we deal with the situation when the difference of  two integrals vanishes, 
but  each integral  does not necessarily vanish.

\subsubsection{``$D$''  part}

For the integral involving the border function, we have 
\begin{align}  &
 P \int_{-1}^1  \frac{H_D (x, x) }{x-\xi}\,  dx 
 =  
 P  \int_{-1}^1  \frac{H_D (x, x) }{x}  \frac{x}{x-\xi}\,  dx
   \nonumber \\ &  =  
P   \int_{-1}^1  dx \,  \frac{x}{x-\xi}\,   \delta (x ) 
 \int_{-1} ^{1 }  \frac{D(\alpha)}{\alpha(1-\alpha)}  
\, d\alpha = 0  \  .
\label{HDxi}
   \end{align}
      In  simple words,   the starting integrand  in (\ref{HDxi}) 
      vanishes for  $x \neq 0$ since   then  $H_{D}(x,x)=0$,
      while    for $x=0$ it is   given   by the $x \delta (x)$
distribution    which  produces zero  after  integration 
with a function that   is  finite   for $x=0$,  which  is  the case
 if  $\xi \neq 0$.     
   Comparing this result  with Eq.~(\ref{eq_SRDDxx}),
   we see that the nonzero value given by the latter  cannot be obtained 
   by  taking $\xi=0$ 
   in the final  result of  Eq.~(\ref{HDxi})  above.
   
  The second piece is given   by 
  \begin{align}  
  &
P  \int_{-1}^1  \frac{ H_D(x,\xi)}{x-\xi}\,  dx  
  =  
  P \int_{-1}^1  \frac{ H_D(x,\xi)}{x}  \,  \frac{x}{x-\xi}\,  dx 
   \nonumber \\   &=   P  \int_{-1}^1  \frac{x\,  dx}{x-\xi}\,    
  \int_{-1} ^{1 }  \frac{D(\alpha)}{\alpha} \,   
    \delta (x - \xi \alpha) \, d\alpha  
   \nonumber \\  & =
  \int_{-1} ^{1 }  \frac{ \xi \alpha }{\xi \alpha - \xi}   \frac{D(\alpha)}{\alpha} \, \, d\alpha   
 = - \int_{-1}^1  \frac{D(\alpha)}{1-\alpha} \, d \alpha  \ .
 \label{SRD2nd}
   \end{align}
   Again,   the result above may be obtained by simply using
   $H_{D}(x,\xi)={\rm sign}(\xi)\,  \theta (|x|<|\xi|)
   D(x/\xi)$ and rescaling \mbox{$x=\alpha \xi$. } 
   Also, 
  though the   final  result of Eq.~(\ref{SRD2nd})   does  not   depend on $\xi$,  
  it   does  not   coincide with  the 
   result of the counterpart relation (\ref{Hdxi1}).
   
 However,  for the difference of  the two  integrals we obtain
\begin{align} 
 & P \int_{-1}^1  \frac{H_D (x, x)}{x-\xi}  \,  dx  -
  P \int_{-1}^1  \frac{H_D(x,\xi)}{x-\xi}  \,  dx= \int_{-1}^1  \frac{D(\alpha)}{1-\alpha} \, d \alpha  \ ,
 \label{eq_SRG3} 
   \end{align}
the same result as in Eq.~(\ref{eq_SRDD}).
Combining the results for the ``plus'' and $D$-parts gives Eq.~(\ref{eq_SRG}).
 
 \subsubsection{Some conclusions}
 
 Thus, our  calculation confirms  the generic GPD sum rule (\ref{eq_SRG})
 derived  in Refs.~\cite{Anikin:2007yh,Diehl:2007jb}. 
 We  were also able  to derive the  $\xi=0$ sum rule
  (\ref{eq_SR})  suggested in Ref.~\cite{Anikin:2007yh}.
It should  be emphasized that the integrals  present 
in the generic sum  rule have  a singularity
  for $x=\xi$,  which is  inside the region of integration,  so the integrals
   may be taken using the principal value prescription.
  Since $H(x,0)/x$ and $H(x,x)/x$ are even  functions
  of $x$, the \mbox{$\xi=0$}  sum rule may be written through  an integral 
  from 0 to 1, and its  $1/x$ singularity is 
  at the end-point of the integration region,
  which means that the $P$-prescription  cannot regulate it.
   Just because of this fact  alone, the sum rule   (\ref{eq_SR}) 
   cannot be a straightforward  consequence of the generic sum rule  (\ref{eq_SRG}).
   
     In  our  derivation, we managed to obtain  finite  expressions
      for each  term involved.  In particular, 
      we   established that though  $H_{D}(x,x)$ and $H_{D}(x,\xi)$
   contributions to the generic sum rule (\ref{eq_SRG})
   are  $\xi$-independent, they do not coincide with 
   their counterparts from the secondary  sum  rule 
     (\ref{eq_SR}), i.e., the latter cannot be obtained 
     by formally continuing to $\xi=0$  the   $\xi$-independent  results 
     for each term of  the generic GPD sum rule. 
   
 In  our derivation, we did not  make an assumption 
 about analyticity of the Mellin moments of GPDs. 
We have obtained   GPD sum rules as a   consequence 
 of the polynomiality of GPDs  that follows from Lorentz invariance
 and is  encoded in the DD representation.
 The analyticity  is a much stronger  restriction.
 One   may  try to find   out whether  it  can   be tested experimentally  
 and  it is  also worth trying to  prove it  in QCD.

 \section{Summary} 
 
 In this  paper, we discussed some  basic   aspects of 
building models for GPDs using the factorized
DD Ansatz (FDDA)  within the ``single-DD''  formulation.
The main difficulty in the  implementation of such a construction
is  the necessity  to deal with projection onto 
a  more singular function  $f(\beta)/\beta$ (rather than 
just onto
$f(\beta)$)   in  the forward limit.  
This leads to two problems.
First, one encounters non-integrable 
singularities for $\beta=0$ in the integrals 
producing GPDs in the central region $|x|<|\xi|$.
The difficulty is  exaggerated by necessity 
to consider  forward distributions $f(\beta)$ that
have a singular $\beta^{-a}$  Regge behavior at small $\beta$.
Second, if there are no factors suppressing 
the $\beta \sim 0$ region  for the integration  line corresponding to $x=\xi$,
the  combined $1/\beta^{1+a}$ singularity 
leads to a singular $(x-\xi)^{-a}$  behavior for  GPDs 
in the outer region  $x>\xi$  near the border point $x=\xi$. 
Such  a behavior was found in the model of 
Ref.~\cite{Szczepaniak:2007af}. 

In our analysis,  we found that 
this model  gives 
 the single-DD-type representation for the model GPD,
 and thus  above reasoning is applicable to it.
But we  argued, that  a proper  softening of the 
hadron-quark vertices produces a profile function 
$h_{N}(\beta,\alpha)$ that  results, for  $x=\xi$,  in  the  ${\cal O} (\beta^{N})$
suppression factor securing a finite value of the GPD
$H(x,\xi)$  at the border point.

However, the profile factor has no impact on the combined 
$1/\beta^{1+a}$ 
singularity  on the $\beta=0$   line 
inside the support rhombus,  which one faces  when calculating GPDs 
in the $|x|<|\xi|$ 
region. 
The advantage of the model of Ref.~\cite{Szczepaniak:2007af} 
is that it implants the Regge behavior 
 through 
 a subtracted dispersion relation for the hadron-quark scattering amplitude.
We  found that the subtraction provides the   regularization  necessary
for  the calculation of GPDs in the central region, and 
illustrated the behavior of resulting GPDs in models 
with $N=1$ and $N=2$ profiles. 

We also observed that this   model produces a $D$-term contribution,
despite the fact that it uses only the forward distribution 
as an input.  This $D$-term 
contribution appears  because 
the subtraction generated by the dispersion relation 
differs from the subtraction that converts
the original DD into a ``plus'' distribution $[f(\beta,\alpha)]_{+}$.
The latter, by definition,  
 cannot generate a $D$-term. 
We have  shown that the GPD $H_{+}(x,\xi)$  generated by  the 
$[f(\beta,\alpha)]_{+}$ part of the original DD 
(i.e., GPD $H(x,\xi)$ with the $D$-term contribution $D(x/\xi)$ subtracted)
 has a remarkable    property
that the integral of $H_{+}(x,\xi)/x$   over positive values $0\leq x \leq 1$
vanishes. As a result, $H_{+}(x,\xi)$  must be negative in some part 
of the central region, a feature that is absent 
in previous FDDA models  based on  
 two-DD formulation.

Within the single-DD formalism, it is  very natural to  separate  the 
relevant DD $f(\beta, \alpha)$ into the 
``plus'' part $[f(\beta,\alpha)]_{+}$
and the $D$-term. 
We  demonstrated that this  separation 
can be used to rederive 
the GPD sum rule related to the dispersion
relation for the real part of the DVCS amplitude,
and we also gave a  derivation 
of  another  sum rule proposed as the $\xi \to 0$ 
limit of that  generic sum rule.
Our derivation shows that this ``secondary''  sum rule
is not a straightforward  consequence of the generic one. 
In particular,  the principal value  prescription used 
in the generic sum rule needs to be  substituted by 
another prescription,  like the
``plus'' prescription.  
The ``plus'' prescription,   in fact, is automatically
generated by the  separation of DDs into the ``plus'' part and the \mbox{$D$-term.}
We also demonstrated that the contributions into the two sum rules
generated by the same functions  are not in a 
 one-to-one correspondence.  

Summarizing,  using (intentionally)  simplified  models, 
we developed the basic tools that can   be used 
in building realistic GPD models based on the
factorized DD Ansatz within the single-DD formalism.
Future developments in this direction
should include the extension of the presented  methods
onto the cases with $a>1$  Regge behavior,
which would require an extra subtraction in the dispersion relation,
and building models for nucleons and other targets with a
non-zero spin. 

\acknowledgements

I would like to express my deep gratitude to
A.P. Szczepaniak for numerous
and intensive communications about 
Ref.~\cite{Szczepaniak:2007af}   that  initiated this work.
I   thank  \mbox{D. M{\"u}ller,} M.V. Polyakov, K.M. Semenov-Tian-Shansky and 
O.V. Teryaev   for  stimulating  discussions   of  GPD  sum rules. 
I am  grateful  to 
    I.V. Anikin, I.I. Balitsky, A.V. Belitsky, S.J. Brodsky,  M. Burkardt, 
M. Diehl,  \mbox{M. Guidal,} V. Guzey, C. E. Hyde, C.-R. Ji, \mbox{X. D. Ji,}  \mbox{P. Kroll,}
 S. Liuti, J.A.Miller,
 I.V. Musatov, M.V. Polyakov, A.Sch{\" a}fer, 
 \mbox{M.A. Strikman,}  L.Szymanowski, A.W.Thomas,
 B. C. Tiburzi, M. Vanderhaeghen  and \mbox{C. Weiss}
for many inspiring discussions   and communications 
that we had over the years,
and which eventually influenced this work.

 This paper is authored by Jefferson Science Associates,
LLC under U.S. DOE Contract No. DE-AC05-06OR23177. 
The U.S.
Government retains a non-exclusive, paid-up,
irrevocable, world-wide license to publish or reproduce this
manuscript for U.S. Government purposes.
 
 \bibliographystyle{apsrev4-1.bst}
\bibliography{gpd.bib}

 \end{document}